\documentclass[letterpaper,prl,aps,twocolumn,showpacs,superscriptaddress,nofootinbib]{revtex4-1}
\usepackage{bm,color}
\usepackage{mathptmx}
\usepackage[T1]{fontenc}
\usepackage[utf8]{inputenc}
\usepackage{latexsym}
\usepackage{graphicx} 
\usepackage{amsmath} 
\usepackage{amsfonts} 
\usepackage{braket}
\usepackage{subfigure}
\usepackage{lineno}

\def \be{{\bf e}}

	\newcommand{\bbm}{\begin{pmatrix}}
	\newcommand{\ebm}{\end{pmatrix}}

	\usepackage[normalem]{ulem} 
	\definecolor{mgreen}{RGB}{1,123,0}

\definecolor{Nathanblue}{rgb}{0.,0.24,0.51}

\def\be{\begin{equation}}
\def\ee{\end{equation}}
\def\bs#1{\mathbf{#1}} 

\begin{document}

\title{Measuring quantized circular dichroism in ultracold topological matter} 

\author{Luca Asteria}
\affiliation{Institut für Laserphysik, Universität Hamburg, 22761 Hamburg, Germany}
\author{Duc Thanh Tran}
\affiliation{Center for Nonlinear Phenomena and Complex Systems, Universit\'{e} Libre de Bruxelles, CP 231, Campus Plaine, B-1050 Brussels, Belgium}
\author{Tomoki Ozawa}
\affiliation{Interdisciplinary Theoretical and Mathematical Sciences Program (iTHEMS), RIKEN, Wako, Saitama 351-0198, Japan}
\author{Matthias Tarnowski}
\affiliation{Institut für Laserphysik, Universität Hamburg, 22761 Hamburg, Germany}
\affiliation{The Hamburg Centre for Ultrafast Imaging, 22761 Hamburg, Germany}
\author{Benno S. Rem}
\affiliation{Institut für Laserphysik, Universität Hamburg, 22761 Hamburg, Germany}
\affiliation{The Hamburg Centre for Ultrafast Imaging, 22761 Hamburg, Germany}
\author{Nick Fläschner}
\affiliation{Institut für Laserphysik, Universität Hamburg, 22761 Hamburg, Germany}
\affiliation{The Hamburg Centre for Ultrafast Imaging, 22761 Hamburg, Germany}
\author{Klaus Sengstock}
\email{klaus.sengstock@physnet.uni-hamburg.de}
\affiliation{Institut für Laserphysik, Universität Hamburg, 22761 Hamburg, Germany}
\affiliation{The Hamburg Centre for Ultrafast Imaging, 22761 Hamburg, Germany}
\affiliation{Zentrum für Optische Quantentechnologien, Universität Hamburg, 22761 Hamburg, Germany}
\author{Nathan Goldman}
\affiliation{Center for Nonlinear Phenomena and Complex Systems, Universit\'{e} Libre de Bruxelles, CP 231, Campus Plaine, B-1050 Brussels, Belgium}
\author{Christof Weitenberg}
\affiliation{Institut für Laserphysik, Universität Hamburg, 22761 Hamburg, Germany}
\affiliation{The Hamburg Centre for Ultrafast Imaging, 22761 Hamburg, Germany}

\date{\today}

\maketitle

{\bf 
The topology of two-dimensional materials traditionally manifests itself through the quantization of the Hall conductance, which is revealed in transport measurements~\cite{Thouless1982,Hasan2010,Qi2011}. Recently, it was predicted that topology can also give rise to a quantized spectroscopic response upon subjecting a Chern insulator to a circular drive: Comparing the frequency-integrated depletion rates associated with drives of opposite orientations leads to a quantized response dictated by the topological Chern number of the populated Bloch band~\cite{Tran2017, Tran2018}. Here we experimentally demonstrate this intriguing topological effect for the first time, using ultracold fermionic atoms in topological Floquet bands. In addition, our depletion-rate measurements also provide a first experimental estimation of the Wannier-spread functional, a fundamental geometric property of Bloch bands~\cite{Marzari1997,Ozawa2018}. 
Our results establish topological spectroscopic responses as a versatile probe, which could be applied to access the geometry and topology of many-body quantum systems, such as fractional Chern insulators~\cite{Neupert2015}.}

The discovery of topological states of matter has revolutionized band theory~\cite{Thouless1982,Hasan2010,Qi2011} by revealing the importance of the Bloch eigenstates and their geometric and topological properties, as captured by the Berry curvature~\cite{Xiao2010} and topological Chern numbers~\cite{Thouless1982,Hasan2010,Qi2011}. These geometric band properties are associated with the adiabatic motion within a given Bloch band~\cite{Xiao2010}, and lead to striking effects such as the anomalous quantum Hall effect~\cite{Chang2013}. The topological invariant associated with Bloch bands (e.g. the Chern number) cannot be identified through the simple observation of the bulk energy bands, which can be accessed by spectroscopy. However, by evaluating not only the excitation frequencies, but also the excitation strengths~\cite{Flaschner2018}, geometrical and topological properties become directly accessible via spectroscopy and lead to new topological phenomena~\cite{Tran2017, Tran2018, Ozawa2018}. 
In particular, subjecting a Chern insulator to a circular drive, and comparing the frequency-integrated depletion rates $\Gamma^{\rm int}_{\pm}\!=\!\int_0^{\infty}\Gamma_{\pm} (\omega) \text{d}\omega$ resulting from drives of opposite orientation (or chirality, $\pm$), yields a quantized response~\cite{Tran2017}
\begin{equation}
\Delta\Gamma^{\rm int}_{\pm}/A_{\rm cell}=(\Gamma^{\rm int}_{+}-\Gamma^{\rm int}_{-})/2A_{\rm cell}=(E_{\rm sp}/\hbar)^2 C,\label{main_result}
\end{equation}
which is dictated by the Chern number $C$ of the populated band. Here $A_{\rm cell}$ is the area of the unit cell~\cite{SupMat}, $E_{\rm sp}$ and $\omega$ are the strength and frequency of the circular drive, and $h\!=\!2 \pi\hbar$ is Planck's constant. This quantized circular dichroism is rooted in the explicit probing of all possible interband couplings~\cite{Tran2017,Souza2008}, which are at the origin of the Berry curvature and topological Chern number~\cite{Xiao2010}.

	\begin{figure}[t]
		\includegraphics[width=\linewidth]{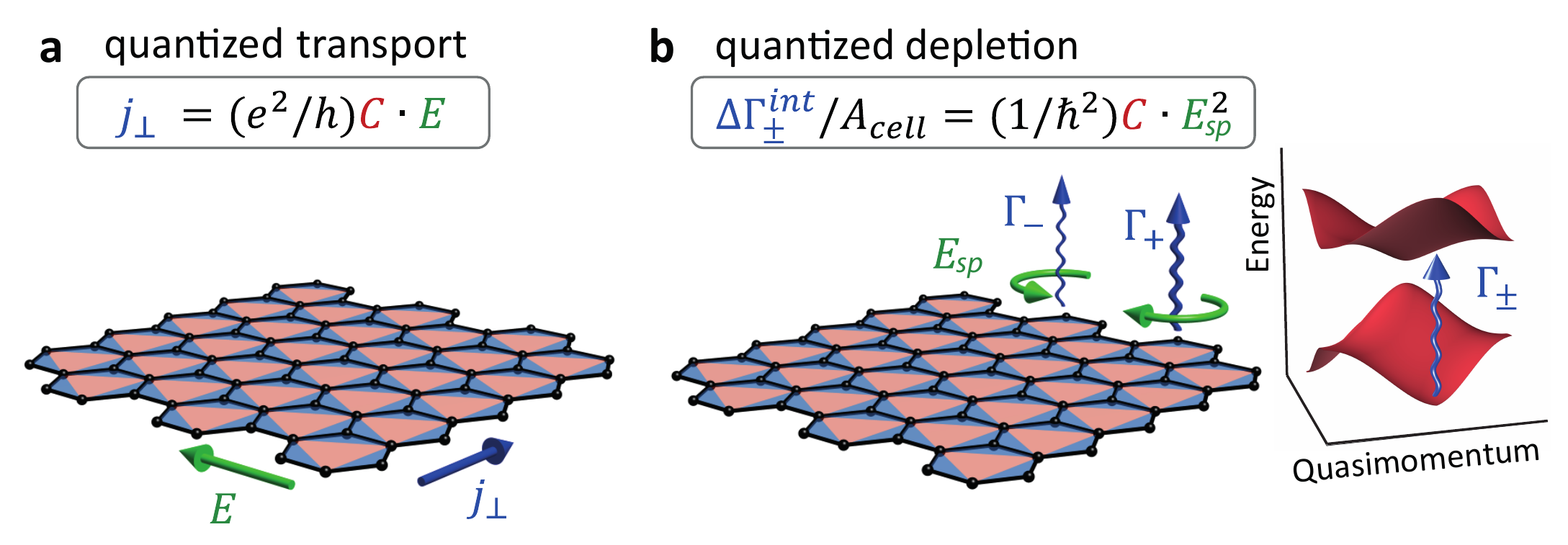}
		\caption{{\bf Quantized responses in topological matter}. {\bf a}, In the quantum (anomalous) Hall effect, the Hall conductance relating the transverse current density $j_{\perp}$ to the applied electric field $E$ follows a quantization law dictated by the Chern number $C$ of the populated Bloch band~\cite{Thouless1982,Xiao2010}. 
		{\bf b}, Our experiment reveals a distinct quantization law~\cite{Tran2017}, which involves the depletion rates $\Gamma_{\pm}$ of a Bloch band (inset) upon circular shaking, where $(\pm)$ refer to the drive orientation. 
		The differential integrated rate $\Delta\Gamma^{\rm int}_{\pm}$ also reveals the Chern number $C$, but is quadratic with respect to the driving strength $E_{\rm sp}$, reflecting its dissipative (interband) nature.}\label{fig:1_Idea}
		\end{figure} 

The quantized circular dichroism predicted in Ref.~\cite{Tran2017} is deeply connected to the better-known quantization of the Hall conductance~\cite{Thouless1982}, $\sigma_H\!=\!(e^2/h)C$, which is observed in transport measurements (Fig.\,\ref{fig:1_Idea}); $e$ denotes the elementary charge. Indeed, the Hall conductivity and the differential depletion rate can be seen as reactive and dissipative responses to an external forcing, respectively, which are connected via the Kramers-Kronig relations~\cite{Bennett1965}. Noting that these quantities are related to the real and imaginary parts of the conductivity tensor, respectively, the quantized circular dichroism in Eq.~\eqref{main_result} can be understood as a simple sum rule for the conductivity tensor~\cite{Bennett1965,Souza2008,Tran2017}. 
A major difference between the (reactive) quantum Hall effect and the quantized dissipative response in Eq.~\eqref{main_result} appears in the different scaling with respect to the driving strength $E_{\rm sp}$:~it is linear in the reactive case while being quadratic in the dissipative case~\cite{Bennett1965}. We note that quantized responses of topological matter have also been identified through other phenomena, such as in the quantization of the Faraday rotation in topological insulators~\cite{Wu2016b} and in the circular photogalvanic effect in Weyl semimetals~\cite{deJuan2017}. While circular dichroism was previously studied in solid-state systems via the irradiation of circularly-polarized light~\cite{Wang2013circular,Sie2015,Gullans2017,Liu2018}, the quantization law predicted in Eq.~\eqref{main_result} has never been observed so far.

	\begin{figure}
		\includegraphics[width=\linewidth]{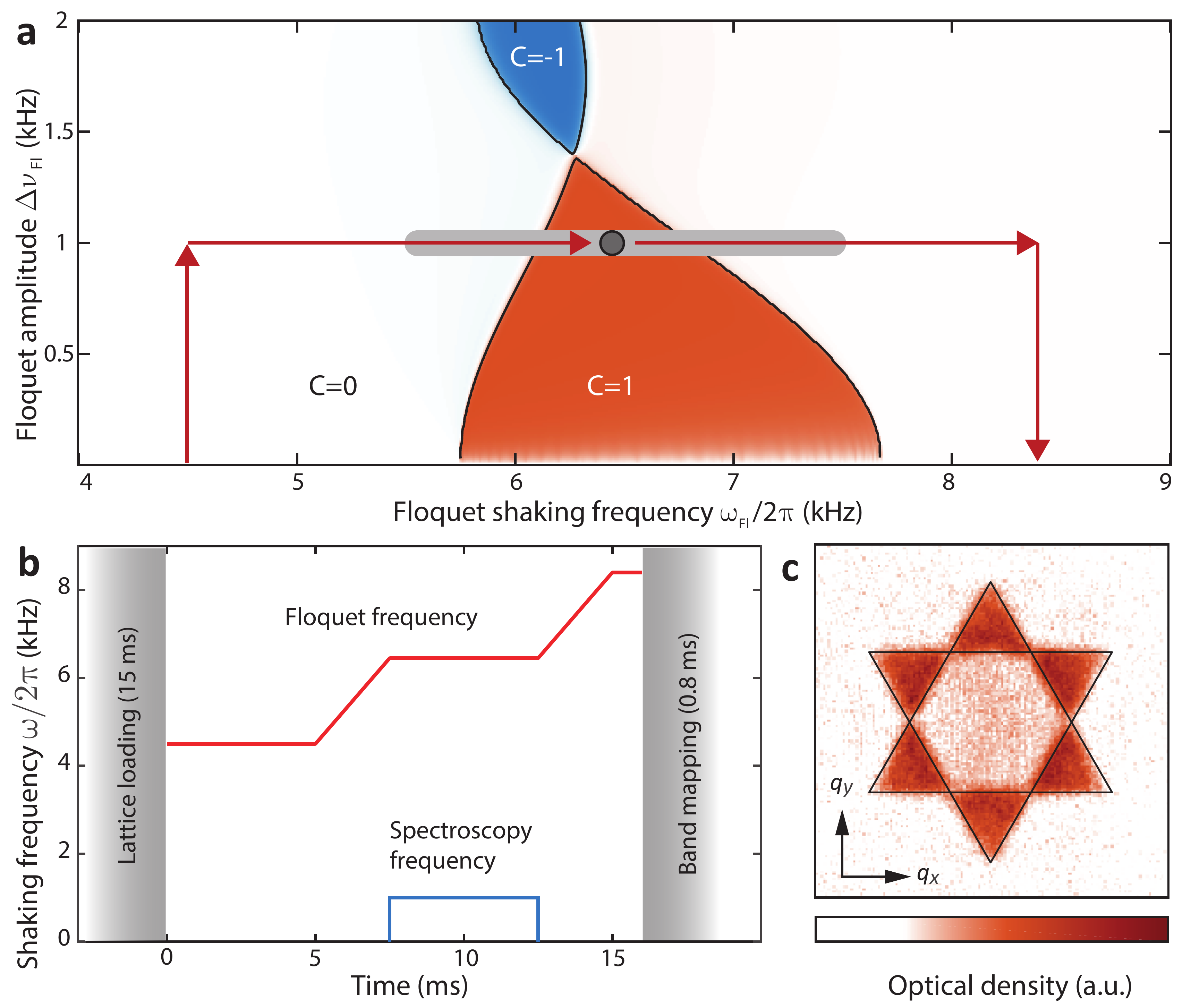}
		\caption{{\bf Measurement scheme}. {\bf a}, Topological phase diagram of the shaken honeycomb lattice, where different regions refer to the Chern number $C$ of the lowest Floquet band. The latter is adiabatically prepared by ramping the Floquet shaking amplitude and frequency (red arrows). After the spectroscopy pulse, applied at varying Floquet frequencies (grey circle), the populations of the two Floquet bands are measured by mapping back onto the bare bands (red arrows) and subsequently performing band mapping. {\bf b}, Timing sequence of the measurement protocol illustrating the separation of time scales between the Floquet frequency and the spectroscopy frequency. {\bf c}, Typical single shot image of the atomic density after band mapping onto quasimomentum ($q_x$ and $q_y$). The relative populations of the first and second bands are obtained by summing up the signal in the first and second Brillouin zone (hexagon and triangles, respectively). Due to the ramping to the other side of the resonance, the lower Floquet band is mapped onto the second Brillouin zone.}\label{fig:2_Scheme}
	\end{figure}

In order to probe the quantized circular dichroism in Eq.~\eqref{main_result}, we consider the Haldane model on a honeycomb lattice~\cite{Haldane1988}. This two-band model, whose bands are topologically non-trivial ($C\!\ne\!0$) for certain parameters, can be realized in ultracold atoms through Floquet-engineering, namely, by circular shaking of a honeycomb optical lattice~\cite{Jotzu2014, Flaschner2016, Tarnowski2017b, Eckardt2017}. The lattice geometry based on three interfering laser beams \cite{Becker2010,Struck2011} is especially well suited, because it allows for fast Floquet driving and therefore a clear separation of time scales.
We start with a sublattice energy offset $\Delta_{AB}\!=\!h\cdot 6.1\,$kHz \cite{Soltan-Panahi2011}, which is large compared to the tunneling element $J\!=\!h\cdot 564\,$Hz, and use near-resonant shaking with Floquet frequency $\omega_{\rm Fl}\!\approx\!\Delta_{AB}/\hbar$ to restore tunneling and create Floquet bands with non-trivial Chern numbers (see Fig.\,\ref{fig:2_Scheme}a for a phase diagram). We employ $^{40}$K atoms in a lattice of wavelength $\lambda\!=\!1064\,$nm, such that the typical length and energy scales are given by the lattice spacing $a_{\rm lat}\!=\!\frac{1}{\sqrt{3}}\frac{2}{3}\lambda\!=\!410\,$nm and the recoil energy $E_{\rm r}\!=\!h\cdot4.41\,$ kHz. 

In our experiment, we probe the properties of a Floquet-engineered system through a spectroscopic measurement, which is based on applying a secondary (perturbative) drive of frequency $\omega_{\rm sp}$. Such an approach is valid whenever a clear separation of time scales exists between the Floquet and spectroscopic drives, $\omega_{\rm Fl}\!\gg\!\omega_{\rm sp}$. In our setting, this separation is obtained through the choice of a large energy offset $\Delta_{AB}\!\approx \hbar\omega_{\rm Fl}\!\gg\!J$, noting that the tunneling $J\!\sim\!\hbar\omega_{\rm sp}$ sets the energy scale of the probed Floquet bands. We have validated this approach through a numerical study of the interplay between the two drives, in particular, the role of micromotion~\cite{SupMat}. We find that the separation of time scales indeed suppresses micro-motion effects and that, for our parameters, they can lead to an error of the integrated depletion rates of about $10\%$. While the spectroscopy drive commutes with the Floquet drive in our experiment, our numerical analysis also indicates that special care is required whenever this commutation relation is not satisfied, which could be the case in other Floquet-based settings. In particular, we find that additional effects, such as a renormalization of the observed coupling strengths, can appear in these situations~\cite{SupMat}.

Our measurements start with spin-polarized fermions filling up the lowest band of the bare lattice. We adiabatically prepare the atoms in the lowest Floquet band by ramping the Floquet amplitude and frequency (Fig.~\ref{fig:2_Scheme}). We then resonantly couple the two lowest Floquet bands via a spectroscopy pulse realized by an additional lattice shaking, which is either circular (with chirality $+$ or $-$) or for comparison linear (along the $x$ or $y$ directions). The spectroscopy drive has varying frequency $\omega_{\rm sp}$ and forcing amplitude $E_{\rm sp}$ \cite{SupMat}. We monitor the transfer between the Floquet bands by adiabatically mapping the Floquet bands onto the bare bands, and subsequently onto the first and second Brillouin zones using adiabatic band mapping~\cite{Aidelsburger2015}. The populations in the Floquet bands $\eta_{1,2}\!=\!N_{1,2}/[N_1\!+\!N_2]$ are obtained by counting the atoms $N_1$ and $N_2$ in the respective Brillouin zones (Fig.~\ref{fig:2_Scheme}).  
We note that the experimental depletion rates $\Gamma$ reflect the change in the fractional population ${\rm d}\eta_{1}/{\rm d}t$ rather than a change in the total atom number (as in Ref.~\cite{Tran2017,Tran2018,Ozawa2018}). This motivates the normalization used in Eq.\,(\ref{main_result}) in terms of $A_{\rm cell}$, instead of the 2D system size $A_{\rm syst}$~\cite{SupMat}. 
 
\begin{figure}
		\includegraphics[width=0.9\linewidth]{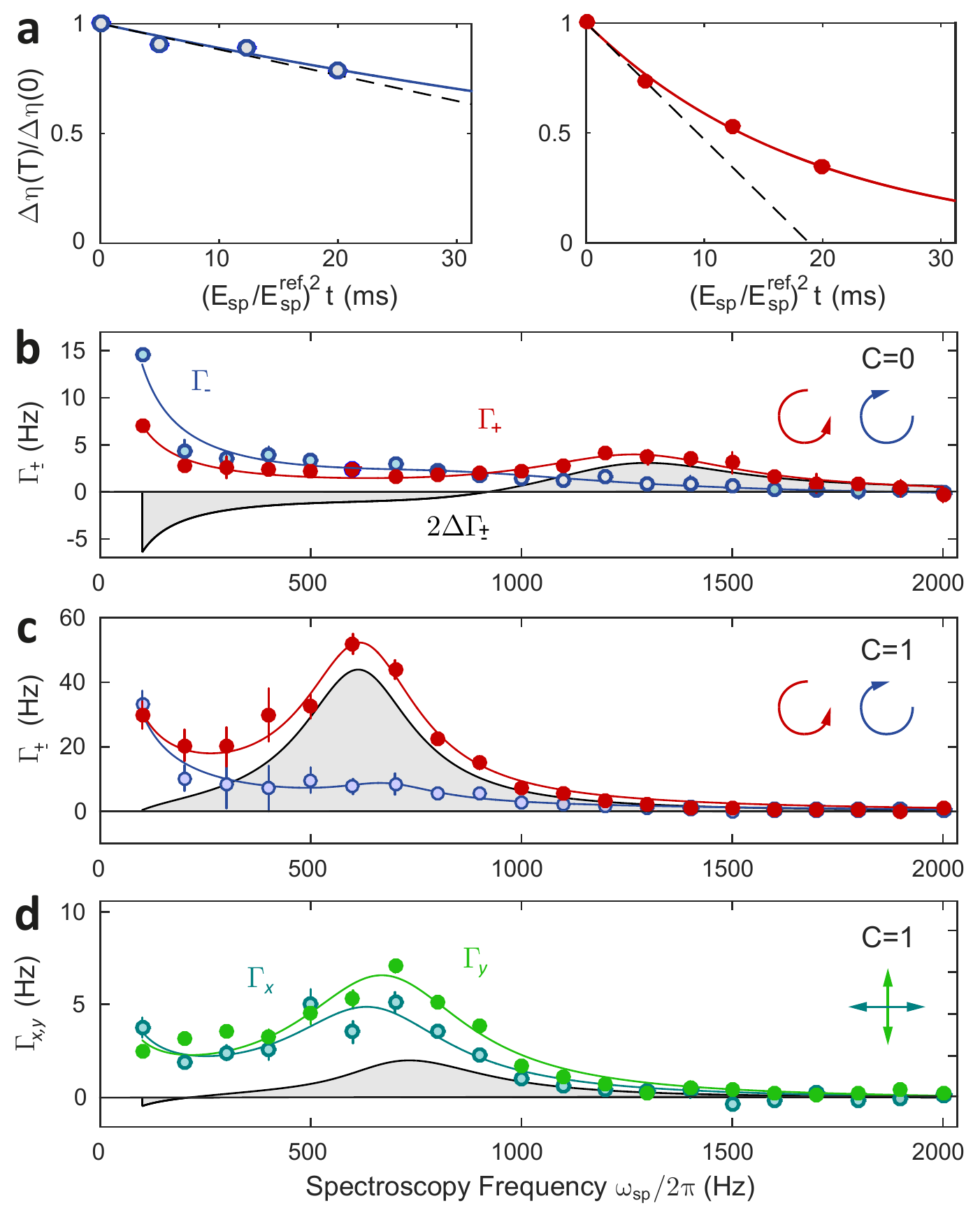}
		\caption{{\bf Chiral spectra of the Floquet bands}. {\bf a}, Exemplary depletion of the lowest band as a function of $(E_{\rm sp}/E_{\rm sp}^{\rm ref})^2 t$ for $t\!=\!5\,$ms and $E_{\rm sp}^{\rm ref}\!=\!0.006 E_{\rm r}/a_{\rm lat}$. The system approaches an equal population of the two bands. We fit an exponential decay $\exp(-2\Gamma (E_{\rm sp}/E_{\rm sp}^{\rm ref})^2 t)$ (solid line) and extract the depletion rate $\Gamma$ at $E_{\rm sp}^{\rm ref}$ in the linear response regime (dashed line). The depletion rate is different for negative chirality (left) and positive chirality (right) of the probe.
		{\bf b}, {\bf c}, Chiral depletion rates $\Gamma_{\pm}$ as a function of the spectroscopy frequency in the $C\!=\!0$ region ($\omega_{\rm Fl}=2\pi\cdot 7.47$\,kHz)(b) and the $C\!=\!1$ region ($\omega_{\rm Fl}\!=\!2\pi\cdot 6.57$\,kHz)(c). The spectra for shaking with positive chirality (red closed circles) and negative chirality (blue open circles) are markedly different, which signals circular dichroism. The lines are a heuristic fit to a Lorentzian peak plus a $1/\omega_{\rm sp}$ term. The differential rate $2\Delta\Gamma_{\pm}$ is indicated by the difference of the two fits (grey area). The y-error bars show the error from the fit to the exponential decay. 
		{\bf d}, Depletion rates $\Gamma_{x,y}$ resulting from linear shaking along the $x$ direction (closed light green circles) and $y$ direction (open dark green circles) in the $C\!=\!1$ region ($\omega_{\rm Fl}\!=\!2\pi\cdot 6.45$\,kHz). The difference in the two rates only shows a very weak linear dichroism (grey area).}\label{fig:3_ChiralSpectra}
	\end{figure}

The spectroscopy drive couples the two Floquet bands and leads to a depletion of the lower band, characterized by the rate $\Gamma$. We start with an initial population fraction in the lower band of $\eta_1(0)\!=\!60-75\%$, due to the compromise between adiabaticity and Floquet heating during the preparation ramp~\cite{Weinberg2015}. Due to dephasing in the inhomogeneous system, the population of the upper band can be described as completely incoherent~\cite{Aidelsburger2015}; therefore, the excess population of the lower band $\Delta \eta(t)\!=\!\eta_1(t)\!-\!\eta_2(t)$ decays to zero with time $t$. We fit $\Delta \eta(t)/\Delta \eta(0)$ with an exponential decay $\exp(-2\Gamma (E_{\rm sp}/E_{\rm sp}^{\rm ref})^2 t)$, and we obtain the desired depletion rate $\Gamma$ at a spectroscopy amplitude of $E_{\rm sp}^{\rm ref}\!=\!0.006\,E_{\rm r}/a_{\rm lat}$ \cite{SupMat}; see Fig.\,\ref{fig:3_ChiralSpectra}a. We choose a fixed duration for the rectangular spectroscopy pulse of $t\!=\!5\,$ms, as a compromise between frequency resolution and pulse duration, and we vary the spectroscopy amplitude $E_{\rm sp}$. In order to obtain a good signal to noise ratio, we probe at driving amplitudes beyond linear response. However, the procedure of fitting the exponential decay allows one to obtain the slope in the linear response regime, described by Fermi's golden rule~\cite{Tran2017} (dashed line in Fig.\,\ref{fig:3_ChiralSpectra}a).

The spectra $\Gamma_{\pm}(\omega_{\rm sp})$ resulting from the circular-driving probe are shown in Fig.\,\ref{fig:3_ChiralSpectra}b, c. As expected, these spectra strongly depend on the chirality of the spectroscopy drive. Specifically, the overall signal is larger when the Floquet and spectroscopy drives have opposite chirality (the Floquet drive has negative chirality throughout the manuscript). These absorption spectra offer a unique characterization of our Floquet band structure and signal the chiral (time-reversal-symmetry-breaking) nature of the engineered Haldane model~\cite{Haldane1988}. We fit the spectra with a heuristic function, composed of a Lorentzian peak and a $1/\omega_{\rm sp}$ term, which captures an additional heating feature at low frequencies. We attribute this heating to the initial jump in the lattice velocity upon switching on/off the spectroscopy pulse; it is independent of the chirality of the spectroscopy drive and hence does not affect the topological response. Introducing the differential rate $\Delta\Gamma_{\pm}\!=\!(\Gamma_+ - \Gamma_-)/2$, our data in Fig~\ref{fig:3_ChiralSpectra}b, c can also be interpreted as a measurement of the dissipative (i.e. imaginary) part of the antisymmetric optical conductivity~\cite{Bennett1965,Tran2017}, $\sigma_{I}^{xy}(\omega)\!=\!\hbar\omega \Delta\Gamma_{\pm}(\omega)/4 A_{\rm cell}E_{\rm sp}^2$. We note that the optical conductivity of a neutral gas~\cite{Wu2015} was also recently measured in a non-topological system~\cite{Anderson2017}.

	\begin{figure}
		\includegraphics[width=\linewidth]{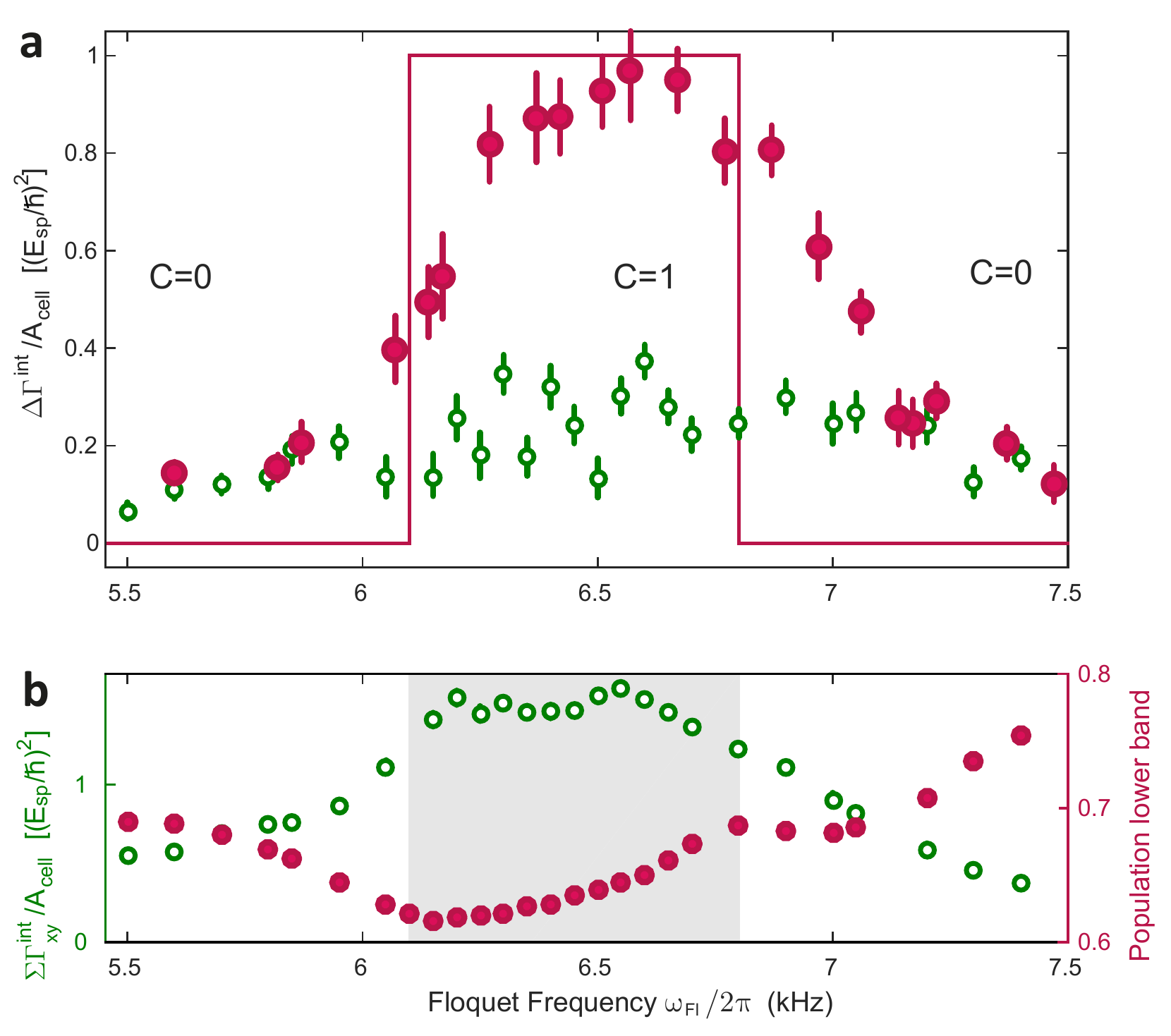}
		\caption{{\bf Spectroscopic signals across the topological phase diagram}. {\bf a}, Circular dichroism measured via $\Delta\Gamma^{\rm int}_{\pm}/A_{\rm cell}$ (closed magenta circles) in units of $(E_{\rm sp}/\hbar)^2$ as a function of the Floquet frequency. The signal is obtained by integrating the Lorentzian parts of the fits to the differential rate spectra in Fig.\,\ref{fig:3_ChiralSpectra}.  The error bars denote the $\chi^2$ of the fit. The signal reaches $0.9(1)$, which is close to the predicted value of $1$ (magenta line), in the center of the $C\!=\!1$ region. In contrast, the linear dichroism $\Delta\Gamma^{\rm int}_{xy}/A_{\rm cell}$ (open green circles) is constant and weak, in the range of $0.2(1)$. {\bf b}, Linear spectroscopic signal $\Sigma\Gamma^{\rm int}_{xy}/A_{\rm cell}$ (open green circles) measured via the sum of spectra resulting from linear shaking, as a function of the Floquet frequency. The signal provides an estimation of the Wannier-spread functional. The signal $\Sigma\Gamma^{\rm int}_{xy}/A_{\rm cell}$, as well as the initial population in the lowest band $\eta_1(0)$ (closed magenta circles), show a change of behavior across the topologically non-trivial region. The grey area (6.1\,kHz - 6.8\,kHz) marks the region with $C=1$ as obtained from a numerical calculation for the independently-determined lattice parameters.}\label{fig:4_Dichroism}
	\end{figure}

The topological response in Eq.~\eqref{main_result} becomes visible when evaluating the frequency-integrated differential rate $\Delta\Gamma^{\rm int}_{\pm}=\int d\omega_{\rm sp} [\Gamma_+(\omega_{\rm sp})-\Gamma_-(\omega_{\rm sp})]/2$; this integration is taken over the Lorentzian part of the fit to the spectra. Fig.\,\ref{fig:4_Dichroism}a shows $\Delta\Gamma^{\rm int}_{\pm}/A_{\rm cell}$ in units of $(E_{\rm sp}/\hbar)^2$ across the topological phase transition (with $A_{\rm cell}\!=\!2\lambda^2/3\sqrt{3}$). As a central and fascinating result of these studies, the signal reaches the value $0.9(1)$, very close to the predicted value of $1$, in the center of the topological ($C\!=\!1)$ region, while a value of $0.2(1)$ is found in the trivial ($C\!=\!0$) regions. Our measurements thus constitute the first experimental observation of quantized circular dichroism. At the topological phase transitions, the signal is found to drop smoothly from the quantized value; we attribute the absence of a sharp jump to our inhomogeneous system and to Fourier broadening of 200\,Hz of our spectra \cite{SupMat}. 
Besides, the predicted detrimental effect of edge states~\cite{Tran2017} does not seem to contribute to our dichroic signal, which we attribute to the spatial separation of bulk and edge states in our harmonically-trapped system, and to an attenuation of the edge-states signal inherent to our band-mapping technique~\cite{SupMat}.

We also analyze linear dichroism, by measuring the depletion rates associated with a linear shaking of the lattice (Fig.~\ref{fig:3_ChiralSpectra}d). In contrast with the circular dichroism discussed above, the differential integrated rate $\Delta\Gamma^{\rm int}_{xy}\!=\!\int d\omega_{\rm sp} [\Gamma_y(\omega_{\rm sp})\!-\!\Gamma_x(\omega_{\rm sp})]/2$ remains approximately constant across the phase transition (Fig.\,\ref{fig:4_Dichroism}a), in the range $0.2(1)$. More importantly, the sum of these integrated rates, $\Sigma\Gamma^{\rm int}_{xy}\!=\!\int d\omega_{\rm sp} [\Gamma_x(\omega_{\rm sp})\!+\!\Gamma_y(\omega_{\rm sp})]/2$, gives access to the Wannier-spread functional $\Omega_I$~\cite{Marzari1997,Ozawa2018}, a fundamental geometric property of Bloch bands that sets a lower bound on the quadratic spread of the underlying Wannier functions [note that this lower bound is only relevant in the trivial ($C\!=\!0$) regime]. As shown in Ref.~\cite{Ozawa2018}, the gauge-invariant part of the Wannier-spread functional is given by 
\begin{equation}
\Omega_I=(\hbar/E_{\rm sp})^2(1/2\pi)\Sigma\Gamma^{\rm int}_{xy}. \label{eq_Wannier}
\end{equation}
In Fig.\,\ref{fig:4_Dichroism}b, we show the spectroscopic signal $\Sigma\Gamma^{\rm int}_{xy}/A_{\rm cell}$ in units of $(E_{\rm sp}/\hbar)^2$, which can be interpreted as a measurement of $\Omega_I$ in units of $A_{\rm cell}/2\pi$~\cite{SupMat}. As predicted~\cite{Ozawa2018}, the signal increases in size upon approaching the phase transitions from the trivial regime. The value of $\Omega_I\approx A_{\rm cell}/2\pi$ close to the phase transition corresponds to a linear extension of the Wannier function on the order of 300\,nm. We note that our signal is expected to exhibit peaks at the phase transitions~\cite{Ozawa2018}. However, these are found to be slightly displaced within the topological region.
We point out that this data constitutes the first experimental estimation of the Wannier-spread functional, a geometric property rooted in the quantum metric tensor~\cite{Marzari1997}. 

We also plot the initial population of the lower band $\eta_1(0)$, which provides a measure for the adiabatic preparation of the Floquet bands. The preparation with fixed ramp time is least successful when targeting states located close to the topological phase transitions, where the gap closing prohibits adiabaticity. The behavior of the initial population $\eta_1(0)$ and of the linear spectroscopic signal $\Sigma\Gamma^{\rm int}_{xy}$ across the phase transition both offer complementary probes for the phase diagram in Fig.~\ref{fig:4_Dichroism}.

Our measurements confirm the quantization of circular dichroism in topological matter, and establish depletion-rate measurements as a versatile probe for the geometric and topological properties of quantum systems. A fully momentum-resolved spectroscopy would allow accessing the Berry curvature~\cite{Tran2017} and quantum metric tensor~\cite{Ozawa2018} in ultracold atoms. Besides, time-resolved chiral spectroscopy could be used to reveal the out-of-equilibrium evolution of topological states under a quench~\cite{Schuler2017}. Finally, the quantized circular dichroism revealed in this work could be generalized in view of probing the topological order of strongly-interacting systems, such as the fractional nature of the Hall conductivity in fractional Chern insulators~\cite{Neupert2015}. 

\begin{acknowledgments}
The authors acknowledge insightful discussions with N.~R. Cooper, M. Dalmonte, A.~G. Grushin, C. Repellin and thank P. Zoller for a careful reading of the manuscript. We acknowledge financial support from the Deutsche Forschungsgemeinschaft via the Research Unit FOR 2414 and the excellence cluster “The Hamburg Centre for Ultrafast Imaging - Structure, Dynamics and Control of Matter at the Atomic Scale”. BSR acknowledges financial support from the European Commission (Marie Curie Fellowship). Work in Brussels is supported by the FRS-FNRS (Belgium) and the ERC Starting Grant TopoCold. TO is supported by the Interdisciplinary Theoretical and Mathematical Sciences Program (iTHEMS) at RIKEN.
\end{acknowledgments}


\bibliographystyle{naturemag}

\newpage
\appendix
\section{SUPPLEMENTAL MATERIAL}

\section{Model for depletion rate with initial occupation}
We derive the model used to describe the depletion rates, which assumes that the population in the upper band can be treated as incoherent. We write the rate equations for the depletion of the respective bands as
\begin{equation}
\begin{aligned}
\dot{\eta_1}=-\Gamma(\eta_1-\eta_2)\\
\dot{\eta_2}=-\Gamma(\eta_2-\eta_1).
\end{aligned}
\end{equation}
The population difference $\Delta\eta=\eta_1-\eta_2$ then obeys the rate equation $\Delta\dot{\eta}=-2\Gamma\cdot\Delta\eta$, which is solved by 
\begin{equation}
\frac{\Delta \eta(t)}{\Delta \eta(0)}=\exp(-2\Gamma t).
\end{equation}

\section{Derivation of the forcing amplitude}
The honeycomb lattice is formed by the interference of three lattice beams with wave vectors ${\bf k}_1=k_{\rm lat}(1, 0)$, ${\bf k}_2=k_{\rm lat}(-1/2,\sqrt{3}/2)$ and ${\bf k}_3=k_{\rm lat}(-1/2,-\sqrt{3}/2)$ with $k_{\rm lat}=2\pi/\lambda$. The reciprocal lattice vectors are ${\bf b}_1={\bf k}_1-{\bf k}_2$ and ${\bf b}_2={\bf k}_3-{\bf k}_2$ with length $\delta k=|{\bf b}_1|=\sqrt{3}k_{\rm lat}$. The sublattice offset $\Delta_{AB}$ is realized by polarisation control of the lattice beams. 
A tight binding fit to the measured band structure of the two lowest bands of the static lattice yields the parameters $\Delta_{AB} = h\cdot 6131$\,Hz, $J_{AB} = h \cdot 564$\,Hz, $J_{AA} = h \cdot 123$\,Hz and $J_{BB} = -h \cdot 6$\,Hz (with the definitions as in Ref.\,\cite{Flaschner2016}).

The circular shaking is implemented via frequency modulation of two of the three lattice beams as
\begin{equation}
\delta \nu_{2,3}=2\Delta\nu[\pm \cos(\omega t)+\sqrt{3}\sin(\omega t)], \delta \nu_{1}=0,
\end{equation}
with the shaking amplitude $\Delta\nu$ in units of frequency. The velocity in $y$ direction is given by 
\begin{equation}
v_y=\frac{2\pi(\delta\nu_2-\delta\nu_3)}{\delta k}=\frac{2\pi 4 \Delta\nu \cos(\omega t)}{\delta k}
\end{equation}
and the corresponding force is
\begin{equation}
F_y=-m\frac{d v_y}{dt}=\frac{m\lambda\omega 4\Delta\nu \sin(\omega t)}{\sqrt{3}}\equiv 2E \sin(\omega t).
\end{equation}
Comparing this expression for the $y$ direction with the total time-dependent Hamiltonian
\begin{equation}
\hat{H}_{\pm}(t)=\hat{H}_0+2E[\cos(\omega t)\hat{x} \pm \sin(\omega t)\hat{y}]
\end{equation}
as defined in Ref.\,\cite{Tran2017}, the forcing amplitude is given by 
\begin{equation}
E=\frac{m\lambda\omega 4\Delta\nu}{2\sqrt{3}}=3 m a_{\rm lat} \omega \Delta \nu,
\end{equation}
with the nearest neighbor lattice spacing $a_{\rm lat}=\frac{1}{\sqrt{3}}\frac{2}{3}\lambda=410\,$nm. 
In this work, we use forcing amplitudes for the spectroscopy of $E_{\rm sp}=0.006 E_{\rm r}/a_{\rm lat}$, $0.009 E_{\rm r}/a_{\rm lat}$, and $0.012 E_{\rm r}/a_{\rm lat}$. In order to keep the forcing amplitude $E_{\rm sp}$ constant when varying the spectroscopy frequency $\omega_{\rm sp}$, we adjust the shaking amplitude as $\Delta\nu_{\rm sp}\propto 1/\omega_{\rm sp}$. For the linear spectroscopy shaking, we apply the amplitudes as
\begin{equation}
\hat{H}_{x}(t)=\hat{H}_0+2\sqrt{2}E[\cos(\omega t)\hat{x}].
\end{equation}

For the circular Floquet drive, we keep the shaking amplitude constant at $\Delta\nu_{\rm Fl}=1\,$kHz. In our range of Floquet frequencies (5.5\,kHz to 7.5\,kHz), this corresponds to Floquet forcing amplitudes in the range of $E_{\rm Fl}=0.4...0.54 E_{\rm r}/a_{\rm lat}$. Therefore even the largest applied spectroscopy forcing amplitude is at least a factor of 30 smaller than the Floquet forcing amplitudes.

\section{Normalization of depletion rates}
The honeycomb optical lattice forms an array of tubes, which are harmonically confined in the transverse direction and contain around 100 atoms each. While for realizing strongly-correlated phases, a transverse lattice can be added, for studying single-particle physics with non-interacting atoms, this geometry gives the best signal to noise. In order to avoid relying on a precise calibration on the number of atoms per tube, we only consider the relative fractions of atoms in the bands of the 2d lattice, thus integrating over the third direction. Our depletion rates are then measured in fractional occupation per second, rather than atoms per second. This requires a normalization of the depletion rate with the area of a unit cell $A_{\rm cell}$, rather than the area of the system $A_{\rm sys}$ as in the original derivation of Ref.\,\cite{Tran2017}. We have followed this new convention throughout the manuscript.

\section{The Wannier-spread functional and its measurement}
Understanding the spatial localization of Wannier states, defined in a given Bloch band, has played a crucial role in the systematic construction of maximally-localized Wannier functions~\cite{Marzari1997}. Importantly, the quadratic spread of Wannier functions was found to be lower-bounded by a quantity,
\be
\langle r^2 \rangle - \langle r \rangle^2 \ge \Omega_{\text{I}},
\ee
where $\Omega_{\text{I}}$ is the gauge-invariant part of the so-called Wannier-spread functional. The latter quantity is deeply connected to the quantum geometry of Bloch states defined in a given band, and it can be simply expressed in terms of their related quantum metric tensor:
\be
\Omega_{\text{I}} = \frac{A_{\text{cell}}}{(2 \pi)^2} \int_{\text{FBZ}} \text{d} \bs{k} \,  \text{Tr} g (\bs k) = \frac{1}{N_{\text{cell}}} \sum_{\bs k} \text{Tr} g (\bs k),\label{general_omega_wannier}
\ee
where $A_{\text{cell}}$ is the area of the unit cell, $N_{\text{cell}}$ denotes the number of unit cells in the system, $\text{FBZ}$ denotes the first Brillouin zone, and where $\text{Tr} g (\bs k)$ denotes the trace of the quantum metric tensor; the relation in Eq.~\eqref{general_omega_wannier} is given for a 2D system. 

In Ref.~\cite{Ozawa2018}, it was suggested that measuring the excitation rates of a quantum system under a periodic drive could be used as a probe of the quantum metric tensor. Considering the linear spectroscopy and integrated rates introduced in the main text, and using the results of Ref.~\cite{Ozawa2018}, one obtains the relation
\be
\Sigma \Gamma^{\text{int}}_{xy}= (\Gamma^{\text{int}}_{x}+ \Gamma^{\text{int}}_{y})/2=2 \pi \left (\frac{E_{\text{sp}}}{\hbar}\right )^2 \frac{1}{N_{\text{cell}}} \sum_{\bs k} \text{Tr} g (\bs k),\label{OG_relation}
\ee
where $g$ is the quantum metric associated with the eigenstates in the lowest Bloch band. Here, we point out that the experimental depletion rates entering Eq.~\eqref{OG_relation} reflect the change in the fractional population $\text{d}\eta_1/\text{d}t$ rather than a change in the total atom number~\cite{Ozawa2018}, which explains the presence of the factor $1/N_{\text{cell}}$ in Eq.~\eqref{OG_relation}; see also Section ``Normalization of depletion rates" in this Supplementary Material. Note that the formula above uses two conventions that differ from Ref.~\cite{Ozawa2018}, but which cancel out: a factor of two in the definition of $\Sigma \Gamma^{\text{int}}_{xy}$ and a factor of $\sqrt{2}$ in the definition of $E_{\rm sp}$. Combining Eq.~\eqref{OG_relation} with Eq.~\eqref{general_omega_wannier} yields the result given in the main text [Eq.~\eqref{eq_Wannier}],
\be
\Omega_{\text{I}} = \frac{1}{2 \pi} \left(\frac{\hbar}{E_{\text{sp}}}\right)^2  \Sigma \Gamma^{\text{int}}_{xy},
\ee
which relates the sum of integrated rates associated with linear drives of different  directions to the (gauge-invariant part of the) Wannier-spread functional. We remind that the relation between $\Omega_{\text{I}}$ and the localization of Wannier functions is only meaningful in the trivial ($C=0$) regime~\cite{Thonhauser2006}.

\section{Data of chiral spectra and numerical simulation}
\begin{figure*}
		\includegraphics[width=0.95\textwidth]{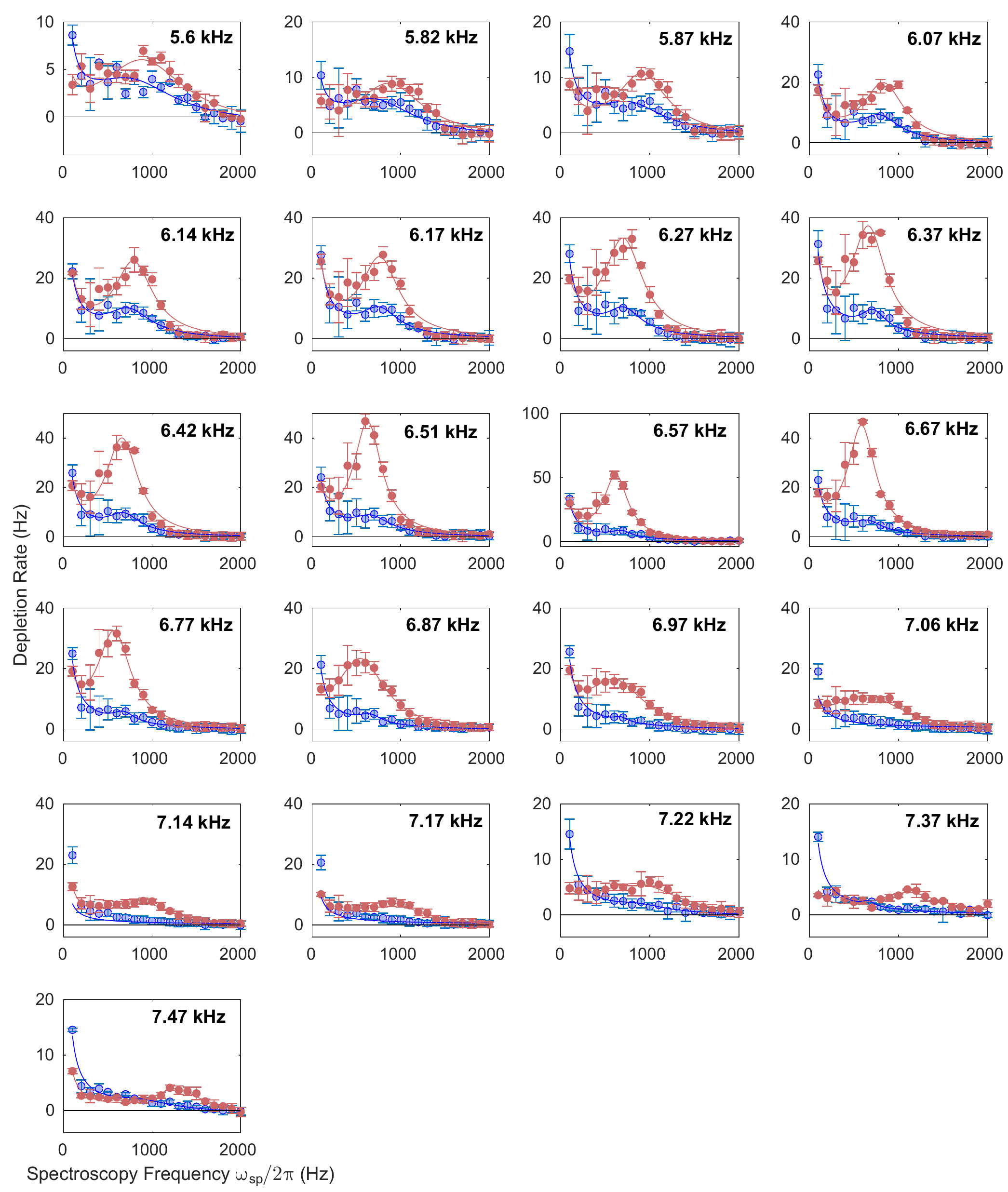}
		\caption{{\bf Experimental spectra for circular shaking}. Depletion rates for probing with positive chirality (red) and negative chirality (blue). The number in the subfigures states the Floquet frequency $\omega_{\rm Fl}/2\pi$.}\label{fig:S1_Spectra}
	\end{figure*}

\begin{figure*}
		\includegraphics[width=0.95\textwidth]{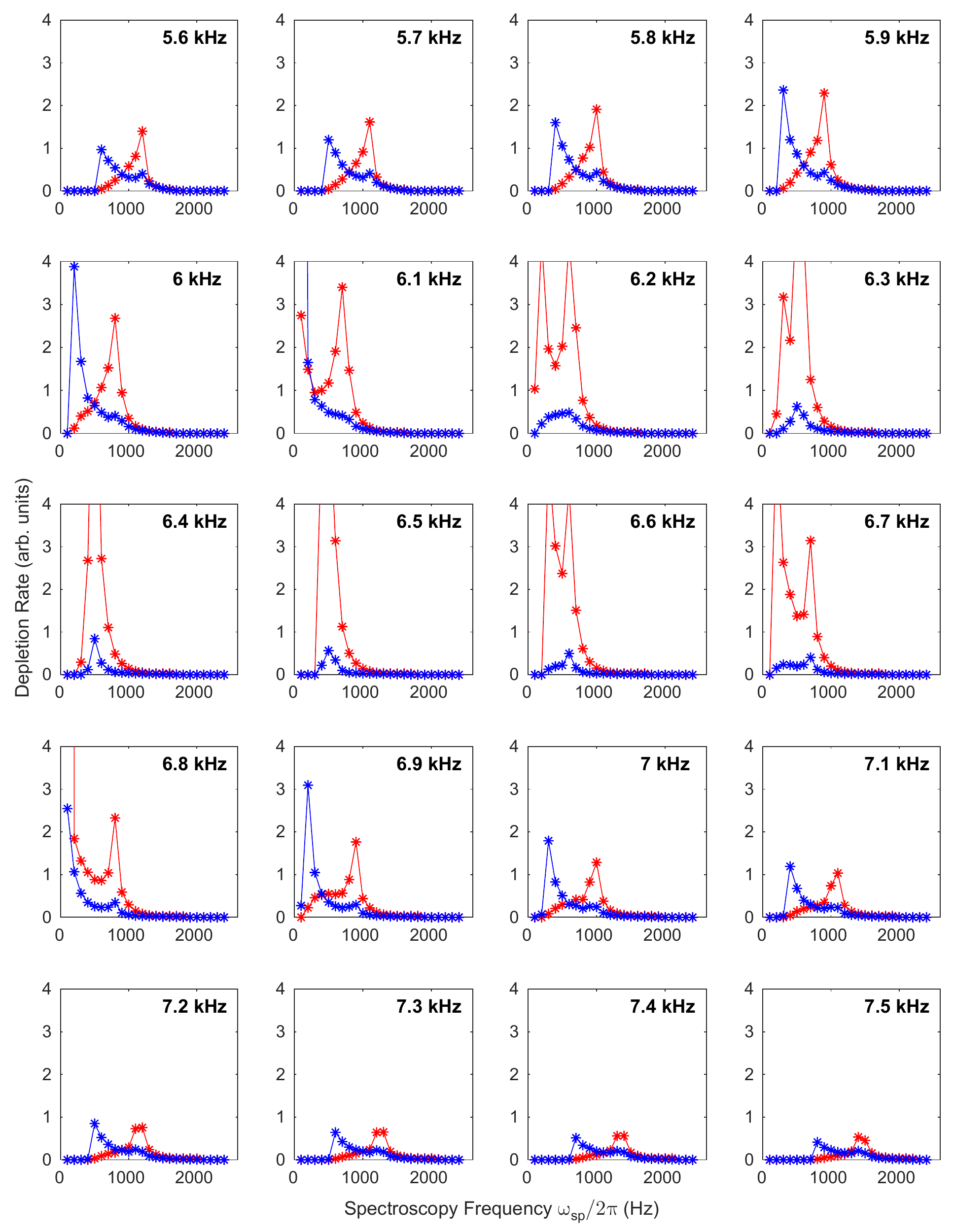}
		\caption{{\bf Numerical spectra for circular shaking}. Depletion rates for probing with positive chirality (red) and negative chirality (blue). The number in the subfigures states the Floquet frequency $\omega_{\rm Fl}/2\pi$.}\label{fig:S2_Spectra}
		
	\end{figure*}

\begin{figure*}
		\includegraphics[width=0.95\textwidth]{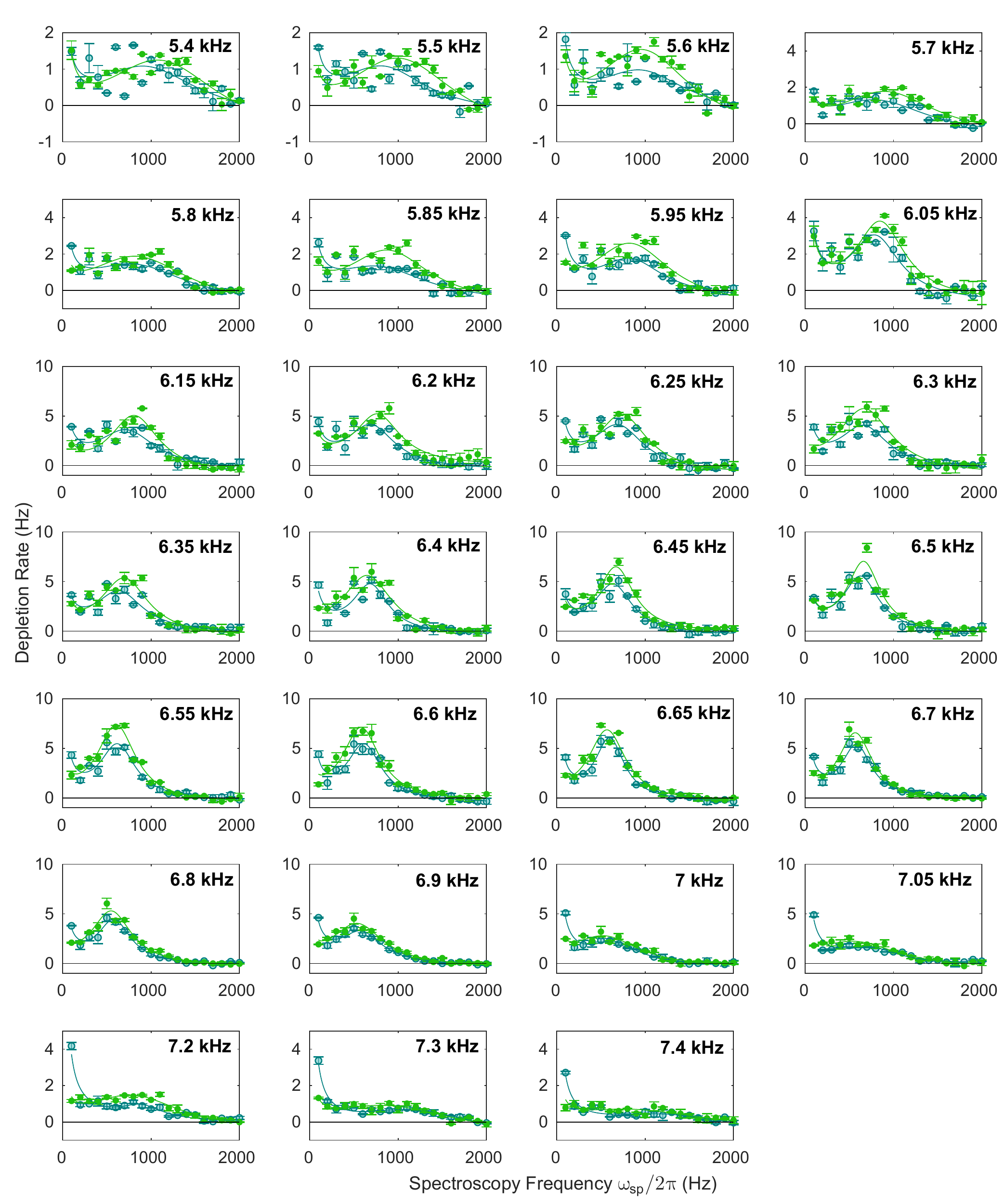}
		\caption{{\bf Experimental spectra for linear shaking}. Depletion rates for probing along the $x$ direction (dark green) and the $y$ direction (light green). The number in the subfigures states the Floquet frequency $\omega_{\rm Fl}/2\pi$.} \label{fig:spectraLinearData}
	\end{figure*}

\begin{figure*}
		\includegraphics[width=0.95\textwidth]{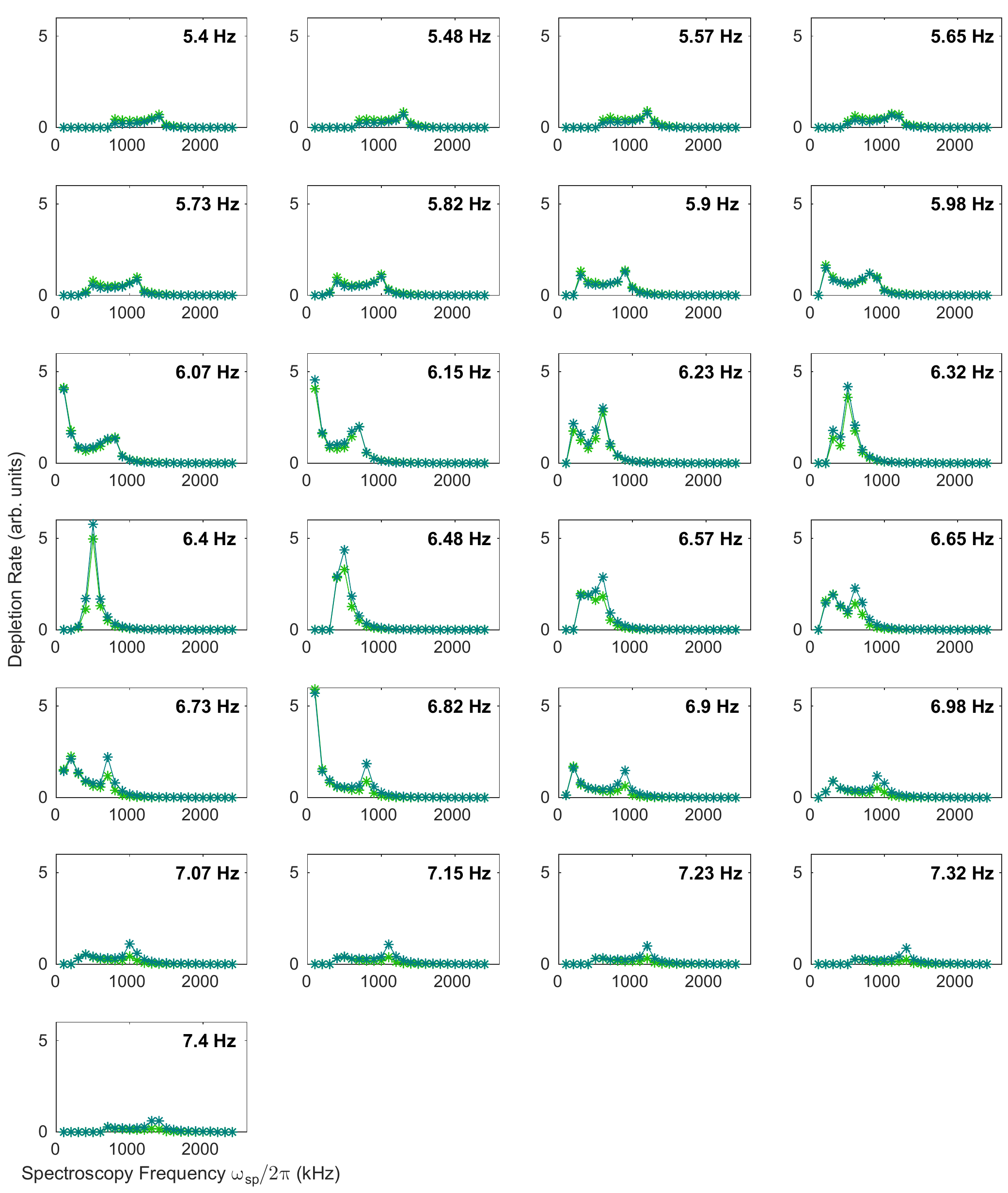}
		\caption{{\bf Numerical spectra for linear shaking}. Depletion rates for probing along the $x$ direction (dark green) and the $y$ direction (light grean). The number in the subfigures states the Floquet frequency $\omega_{\rm Fl}/2\pi$.} \label{fig:spectraLinearNumerics}
	\end{figure*}

In Fig.\,\ref{fig:S1_Spectra} we present the chiral spectra for all Floquet frequencies, which are used to obtain the dichroism shown in Fig.\,\ref{fig:4_Dichroism}.
For comparison, we show numerically obtained spectra for the same parameters in Fig.\,\ref{fig:S2_Spectra}. In Fig.\,\ref{fig:spectraLinearData}, we show the spectra for linear shaking for all Floquet frequencies along with the numerical predictions in Fig.\,\ref{fig:spectraLinearNumerics}. The experimental spectra show the depletion rates at the reference amplitude $E_{\rm sp}^{\rm ref}=0.006\,E_{\rm r}/a_{\rm lat}$. The numerical simulations use a full simulation of the Floquet system (see Supplementary of Ref.~\cite{Flaschner2016}), but do not include the effects of the Fourier width, initial population or external trap.

\newpage
\section{Momentum dependence of the matrix elements}
\begin{figure}
		\includegraphics[width=0.9\linewidth]{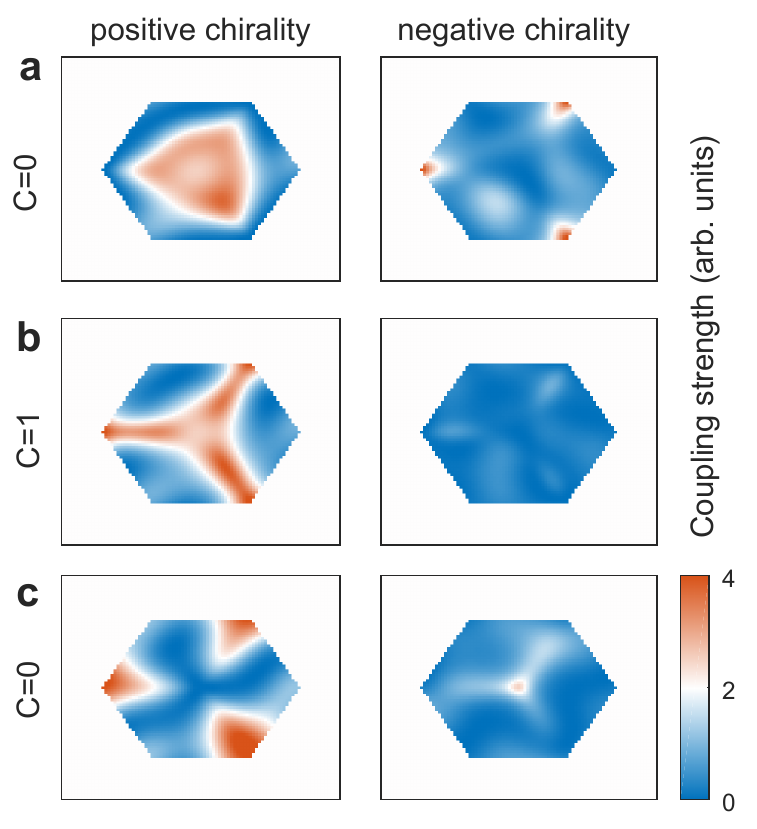}
		\caption{{\bf Numerical matrix elements of the drive between the two Floquet bands}. The matrix elements are evaluated for positive chirality (left) and negative chirality (right) of the perturbation. The Floquet bands are created by driving with negative chirality. The Floquet frequencies are $\omega_{\bf Fl}=2\pi\cdot 5.9$\,kHz (C=0)(a), $\omega_{\bf Fl}=2\pi\cdot 6.6$\,kHz (C=1)(b), and $\omega_{\bf Fl}=2\pi\cdot 7.0$\,kHz (C=0)(c).} \label{fig:matrixElements}
	\end{figure}

The coupling strength of the chiral perturbation between the two Floquet bands is given by the corresponding matrix element. Fig.\,\ref{fig:matrixElements} shows these momentum-resolved matrix elements for both chiralities of the perturbation and for different Floquet frequencies corresponding to $C=0$ and $C=1$ regions. The coupling is very different for the two chiralities, which is - in combination with the band separation and the density of states - at the origin of the circular dichroism. The plots also explain why the depletion spectrum for positive chirality has its weight at higher frequencies than the spectrum for negative chirality for both large and small Floquet frequencies. For red-detuned Floquet frequencies, positive chirality couples most strongly at the $\Gamma$ point, which corresponds here to large band gaps. In contrast, for blue-detuned Floquet frequencies it couples most strongly at the $K$ point, which corresponds to the large band gaps in this regime. The Dirac points at the $\Gamma$ and $K$ point are most relevant for this discussion due to the large density of states at these band extrema. These observations can be related to the phenomenon of valley-selective coupling of circular polarized light in graphene \cite{Yao2008}.
	
\section{Discussion of systematic effects}
An effect that might reduce the dichroism signal is the edge-state contribution, which is expected to completely compensate the bulk response in the ideal system \cite{Tran2017}. This effect would lead to a dichroism of opposite sign at small spectroscopy frequencies, which couple to the edge states. We do not observe such a signal in our frequency-resolved measurements and instead measure the full quantized value for the frequency-integrated differential rate. It seems that edge states do not play an important role for the spectroscopy in our harmonically trapped system. We attribute this to the reduced overlap between the bulk and edge states in the harmonic trap~\cite{Buchhold2012,Goldman2013}, and to an attenuation of the edge-states signal inherent to our band-mapping technique (the edge-states population being most probably redistributed among both Bloch bands). This is also consistent with the quantized Hall drift observed in harmonically-trapped bosonic gases~\cite{Aidelsburger2015}.

The inhomogeneous system and the dynamics induced by the external trap (about 70\,Hz) washes out the momentum-dependence of the excitations \cite{Heinze2013} (see Fig.\,\ref{fig:2_Scheme}{\bf c}). Therefore a momentum-resolved evaluation of the data, which would also reveal the Berry curvature, is not possible. 

We fit the chiral spectra in Fig.\,\ref{fig:3_ChiralSpectra} with a heuristic function composed of a Lorentzian peak and a $1/\omega_{\rm sp}$ term. The heuristic fit function is motivated by the dominant role of Fourier broadening. The $1/\omega_{\rm sp}$ term captures an additional heating feature at low frequencies, which we attribute to the initial jump in the lattice velocity at the switch-on/off of the spectroscopy pulse. This jump necessarily appears in a realization of the circular driving with zero mean velocity and hence zero mean displacement of the lattice after one spectroscopy period. For linear spectroscopy shaking, we choose the initial shaking phase to avoid the jump. The jump in velocity scales as $1/\omega_{\rm sp}$ and is therefore most pronounced at small spectroscopy frequencies of 100\,Hz. 
The initial jump leads to an additional velocity of the atoms in the lattice, 
which in principle can be sensitive to the dependence of the Floquet eigenstates  on quasimomentum. The coefficient of the $1/\omega_{\rm sp}$ term is indeed larger in the non-trivial region, but within statistical errors always equal for the two chiralities.

To obtain the frequency-integrated differential depletion rates, we determine the area under the fits in order to reduce experimental noise. We only use the Lorentzian part of the fit in order to remove the contribution of the $1/\omega_{\rm sp}$ term. As this heating feature is not part of the topological response and independent of the chirality of the spectroscopy drive, it cancels when calculating the differential rate.

\section{Spectroscopy of a Floquet system: \\ Time-scale separation and micro-motion}

In our experiment, we probe the properties of a Floquet-engineered system through a spectroscopic measurement. In this setting, the total system contains two types of time-dependent features: (a) the Floquet drive (`Fl') used to engineer a band structure of interest, and (b) the time-dependent perturbation (`sp') that is used as a  spectroscopic probe. It is the aim of this theory Section to explore the interplay between these two crucial features of our experimental setting. In the following, we set $\hbar\!=\!1$ unless otherwise stated. 

 \paragraph{Basics of Floquet engineering}

Floquet engineering consists in modifying the properties of a system by subjecting it to a time-periodic drive~\cite{Cayssol_review,Goldman_PRX,Eckardt2017}. The corresponding Hamiltonian can thus be written in the general form 
\be
\hat H (t)= \hat H_0 + \hat V_{\text{Fl}}(t), \quad \hat V_{\text{Fl}}(t+T_{\text{Fl}})=\hat V_{\text{Fl}}(t),
\ee
where $\hat H_0 $ is a static Hamiltonian, and where $T_{\text{Fl}}\!=\!2 \pi/\omega_{\text{Fl}}$ is the period of the drive. Since the system is time-periodic, its stroboscopic dynamics -- probed at times multiple of the period $T$ -- can be generated by the time-evolution operator over a single period $\hat U(T)$. Specifically, the state of the system at time $t_N\!=\!N T$ is obtained as $\vert \psi (t_N)\rangle = [\hat U (T)]^N \vert \psi (0)\rangle$, where $\vert \psi (0)\rangle$ denotes the initial state at time $t\!=\!0$. When writing the time-evolution operator in the appealing form $\hat U(T)\!=\!\exp (-i T \hat H_{\text{eff}})$, one can write the time-evolved state as $\vert \psi (t_N)\rangle \!=\! \exp (-i t_N \hat H_{\text{eff}}) \vert \psi (0)\rangle$, which indicates that the system evolves as if it was described by a time-independent effective Hamiltonian $\hat H_{\text{eff}}$. In general, the latter is a complicated operator, which stems from a rich interplay between the static Hamiltonian $\hat H_0$ and the drive $\hat V_{\text{Fl}}(t)$, and it can be explicitly computed using the methods of Refs.~\cite{Goldman_PRX,Goldman_Floquet_PRA,Anisimovas,Mikami}. Importantly, the discussion above builds on a stroboscopic-evolution picture, and in this sense, it disregards the effects of micro-motion (i.e.~the evolution at intermediate times $t_N\!\ne\!N T$). These micro-motion effects can also be systematically studied, using the formalism of Refs.~\cite{Goldman_PRX,Goldman_Floquet_PRA}.

\paragraph{The simplified picture and the hierarchy of scales}

In the experiment, we are interested in probing the properties of an effective Hamiltonian $\hat H_{\text{eff}}$, which results from a Floquet-engineering protocol, as explained above. In analogy with the spectroscopy of static systems, we act on the Floquet-engineered system with an extra time-dependent perturbation, so that the corresponding (total) Hamiltonian can be loosely written as 
\be
\hat{\mathfrak H}_{\text{tot}} (t)\!=\!\hat H_{\text{eff}} + \hat V_{\text{sp}} (t), \quad \hat V_{\text{sp}}(t+T_{\text{sp}})=\hat V_{\text{sp}}(t) , \label{simplification}
\ee
where $\hat V_{\text{sp}} (t)$ describes the effects of the spectroscopic probe, and where we assumed that the Floquet-engineered system can be treated as a static system captured by the effective Hamiltonian $\hat H_{\text{eff}}$. Importantly, the latter simplification should be treated with care, as we will investigate in more detail below. Intuitively, this approach should hold whenever the system presents a clear separation of time-scales:~Introducing the bandwidth of the effective spectrum, $W_{\text{eff}}$, one typically has the hierarchy of scales, 
\be
\omega_{\text{sp}}\lesssim W_{\text{eff}} \ll \omega_{\text{Fl}},\label{hierarchy}
\ee
where $\omega_{\text{sp}}\!=\!2 \pi / T_{\text{sp}}$ denotes the probe frequency. Indeed, Floquet-engineering typically operates in the so-called ``high-frequency regime", where $\hbar \omega_{\text{Fl}}\!\rightarrow\!\infty$ sets the largest energy scale in the problem~\cite{Goldman_PRX,Eckardt2017}. When the hierarchy in Eq.~\eqref{hierarchy} is satisfied, one expects the micro-motion effects to be irrelevant, which justifies the simplification introduced in Eq.~\eqref{simplification}. 

\paragraph{Probing Floquet systems: A numerical study}

It is the aim of this Section to analyze the validity of this probing approach, by studying the full time-dependent Hamiltonian, i.e.~the total Hamiltonian that explicitly contains the two types of time-dependent features, namely,
\be
\hat H_{\text{tot}} (t)= \hat H_0 + \hat V_{\text{Fl}}(t) + \hat V_{\text{sp}} (t),\label{total_ham_gen}
\ee
where both the Floquet drive and the probe are explicitly written, i.e.~without applying the aforementioned simplification [Eq.~\eqref{simplification}]. In our experiment, $\hat H_0$ describes the motion of a particle in a 2D honeycomb lattice with large energy offsets between neighboring sites, while both $\hat V_{\text{Fl}}(t)$ and $\hat V_{\text{sp}}(t)$ describe the effects of circular shaking; we note that the corresponding frequencies $\omega_{\text{sp}}$ and $\omega_{\text{Fl}}$ indeed satisfy the hierarchy in Eq.~\eqref{hierarchy}; see main text. In our setting, the effective Hamiltonian $\hat H_{\text{eff}}$ is reminiscent of the Haldane model~\cite{Haldane1988}, i.e. a two-band lattice model displaying Bloch bands with non-zero Chern numbers. The corresponding Hamiltonian can thus be written in the momentum representation as
\be
\hat H_{\text{eff}} (\bs q)= d_x (\bs q) \hat \sigma_x + d_y (\bs q) \hat\sigma_y + d_z (\bs q) \hat\sigma_z,\label{effective_ham}
\ee
where $\bs{q}$ denotes the quasi-momentum and $\hat \sigma_{x,y,z}$ are the Pauli matrices reflecting the pseudo-spin structure associated with the honeycomb lattice. Following the proposal in Ref.~\cite{Tran2017}, the Chern number of the lowest band $\varepsilon_{-} (\bs q)\!=\! - (d_x^2 + d_y^2 + d_z^2)^{1/2} $, can be probed by subjecting the Haldane model to a circular drive and by measurement the resulting excitation rates (see also the main text); this circular perturbation corresponds to the spectroscopic probe $\hat V_{\text{sp}}(t)$ in Eq.~\eqref{total_ham_gen}. 

When treated in a proper frame~\cite{Tran2017}, the circular drives [$\hat V_{\text{Fl}}(t)$ and $\hat V_{\text{sp}} (t)$] enter the Hamiltonian in Eq.~\eqref{total_ham_gen} as a uniform time-dependent gauge potential that affects the hopping terms in $\hat H_0$; hence, these time-dependent features do not affect the translational symmetry of the problem. Consequently, the total Hamiltonian in Eq.~\eqref{total_ham_gen} still decouples in terms of the quasi-momenta, which results in a set of driven two-level systems (one for each quasi-momentum $\bs q$):
\be
\hat H_{\text{tot}} (\bs q; t)= \epsilon (\bs q) \hat 1 + d_x (\bs q; t) \hat \sigma_x + d_y (\bs q; t) \hat\sigma_y + d_z (\bs q) \hat\sigma_z . \label{total_ham_gen_bis}
\ee

In order to explore the interplay between the different types of time-modulations [$\hat V_{\text{Fl}}(t)$ and $\hat V_{\text{sp}} (t)$], we now consider a similar but simpler model: a time-modulated two-level system described by the Hamiltonian:
\begin{align} 
&\hat H (t)= \hat H_0 + \hat V_{\text{Fl}}(t) + \hat V_{\text{sp}} (t), \label{total_ham}\\
&\hat H_0= \Delta \vert 1 \rangle \langle 1 \vert + J \left (\vert 0\rangle \langle 1\vert + \vert 1\rangle \langle 0\vert \right ),  \notag  \\
&\hat V_{\text{Fl}} (t)= K_{\text{Fl}} \cos (\Delta t) \vert 0 \rangle \langle 0 \vert , \quad  \hat V_{\text{sp}} (t)= 4 K_{\text{sp}} \cos (\omega_{\text{sp}} t) \vert 0 \rangle \langle 0 \vert , \notag 
\end{align}
which is indeed a direct analog of our experimental shaken optical lattice: the two levels (or `sites') are separated by a large offset $\Delta \gg J$, and the effective coupling is reintroduced by the resonant time-modulation $\hat V_{\text{Fl}} (t)$, i.e.~$\omega_{\text{Fl}}\!=\!\Delta$. In the absence of the probe ($K_{\text{sp}}\!=\!0$), the dynamics is well described by the effective Hamiltonian~\cite{Goldman_Floquet_PRA} 
\be
\hat H_{\text{eff}}=J \mathcal{J}_1 (K_{\text{Fl}}/\Delta) \left (\vert 0\rangle \langle 1\vert + \vert 1\rangle \langle 0\vert \right ) + \mathcal{O} (1/\Delta),\label{eff_ham}
\ee
where $\mathcal{J}_1$ is a Bessel function of first kind; this is the basic mechanism for photon-assisted tunneling~\cite{Eckardt2017}. The perturbation $\hat V_{\text{sp}} (t)$ is then introduced to probe the effective system characterized by $\hat H_{\text{eff}}$. Importantly, we note that the operators associated with both the Floquet-drive and the probe commute with each other, $[\hat V_{\text{Fl}},\hat V_{\text{sp}}]\!=\!0$, which is also the case in our experiment (since both effects are generated by circular shaking); this aspect will be further discussed at the end of this Section.

Let us assume for now that the system can be simplified in the `ideal' (or naive) form in Eq.~\eqref{simplification}, namely, let us assume that the Floquet-engineered system can indeed be treated as a static (effective) Hamiltonian, as given in Eq.~\eqref{eff_ham}. For convenience, we now introduce the eigenstates of the effective (Floquet) Hamiltonian:
\begin{align}
\vert + \rangle =(1/\sqrt{2}) \left ( \vert 0 \rangle + \vert 1 \rangle \right ) , \qquad \vert - \rangle =(1/\sqrt{2}) \left ( \vert 0 \rangle - \vert 1 \rangle \right ) ,\notag
\end{align}
which will be referred to as ``dressed states" in the following. Their energies are $\varepsilon_{\pm}\!=\!\pm J \mathcal{J}_1 (K_{\text{Fl}}/\Delta) $, according to Eq.~\eqref{eff_ham}. In this ``dressed-state" basis, the ideal Hamiltonian in Eq.~\eqref{simplification} now explicitly reads
\be
\hat{\mathfrak H}_{\text{tot}} (t)=J \mathcal{J}_1 (K_{\text{Fl}}/\Delta) \hat \sigma_z + 2 K_{\text{sp}} \cos (\omega_{\text{sp}} t) \left ( \hat \sigma_x + \hat 1 \right ).\label{dressed_ham}
\ee
From this, one may propose the following probing protocol: setting the probe frequency on resonance $\omega_s\!=\!2J \mathcal{J}_1 (K_{\text{Fl}}/\Delta)$, and assuming that the lowest-energy dressed-state is initially occupied $\vert \langle - \vert \psi (t\!=\!0)\rangle \vert^2=1$, should result in perfect Rabi oscillations between the two dressed states. Furthermore, extracting the excitation rate $\Gamma$ at short times ($t\!\ll\!1/K_{\text{sp}}$), and integrating over the probe frequency should verify the relation~\cite{Tran2017,Tran2018}
\begin{align}
&\int_0^{\infty} \Gamma (\omega_{\text{sp}}) \text{d} \omega_{\text{sp}} \approx 2 \pi \vert \langle + \vert \hat V_{\text{sp}}^{+}\vert - \rangle \vert^2 = 2 \pi K_{\text{sp}}^2 ,\label{Rabi_integrate}
\end{align}
where $\hat V_{\text{sp}}^{+}$ is defined through $\hat V_{\text{sp}}(t)\!=\!\hat V_{\text{sp}}^{+}e^{-i \omega_{\text{sp}}t} + \hat V_{\text{sp}}^{-}e^{i \omega_{\text{sp}}t}$. We remind that this relation~\eqref{Rabi_integrate} between the integrated excitation rate and the coupling matrix element $\vert \langle + \vert \hat V_{\text{sp}}^{+}\vert - \rangle \vert^2\!$ is at the core of the quantized dichroism effect revealed in the experiment; see Ref.~\cite{Tran2017,Tran2018} for details.

We now numerically explore this protocol, to test the validity of this spectroscopic approach. For practical reasons, we study this scenario in a different frame, as generated by the unitary operator~\cite{Goldman_Floquet_PRA}
\be
\hat R (t)= \exp \left [ i \Delta t \vert 1 \rangle \langle 1 \vert + i (K_{\text{Fl}}/\Delta) \sin (\Delta t) \vert 0 \rangle \langle 0 \vert \right ].\label{frame_change}
\ee
In this frame, the total Hamiltonian in Eq.~\eqref{total_ham} now reads
\begin{align}
\mathcal{H} (t)&=\hat{R} \hat{H}(t) \hat{R}^\dagger - i \hat{R} \partial_t \hat{R}^\dagger \notag \\
&= J \vert 0\rangle \langle 1\vert  \exp \left [ i (K_{\text{Fl}}/\Delta) \sin (\Delta t) -i \Delta t \right ] + \text{h.c.} \notag \\
&+ 4 K_{\text{sp}} \cos (\omega_{\text{sp}} t) \vert 0 \rangle \langle 0 \vert.\label{total_ham_frame}
\end{align}
This is the Hamiltonian that we implement numerically in our time-evolution simulations discussed below. One should note that this Hamiltonian includes all the time-dependences of our problem (i.e. there are no approximation at this level).

Before presenting the numerics, let us discuss the different energy scales. In the following, we set $J\!=\!1$ to be our unit of energy, and we also set $K_{\text{Fl}}/\Delta\!=\!2$ such that $\omega_{\text{sp}}\!=\!2J \mathcal{J}_1 (K_{\text{Fl}}/\Delta)\!\approx\!1$. In this situation, a good separation of energy scales is obtained for $\Delta\!\gg \!1$: the Floquet and probe frequencies satisfy $\omega_{\text{Fl}}\!\gg\!\omega_{\text{sp}}$. Hence, deviations from the ideal situation [Eq.~\eqref{simplification}] are expected for $\Delta \sim 1$. In the following, we also set the perturbation strength $K_{\text{sp}}\!=\!0.01J$ in order to have relatively long Rabi periods.

\begin{figure*}[htb]
\begin{center}
\subfigure[$\Delta = 100J$]{
\includegraphics[width= 0.23 \textwidth]{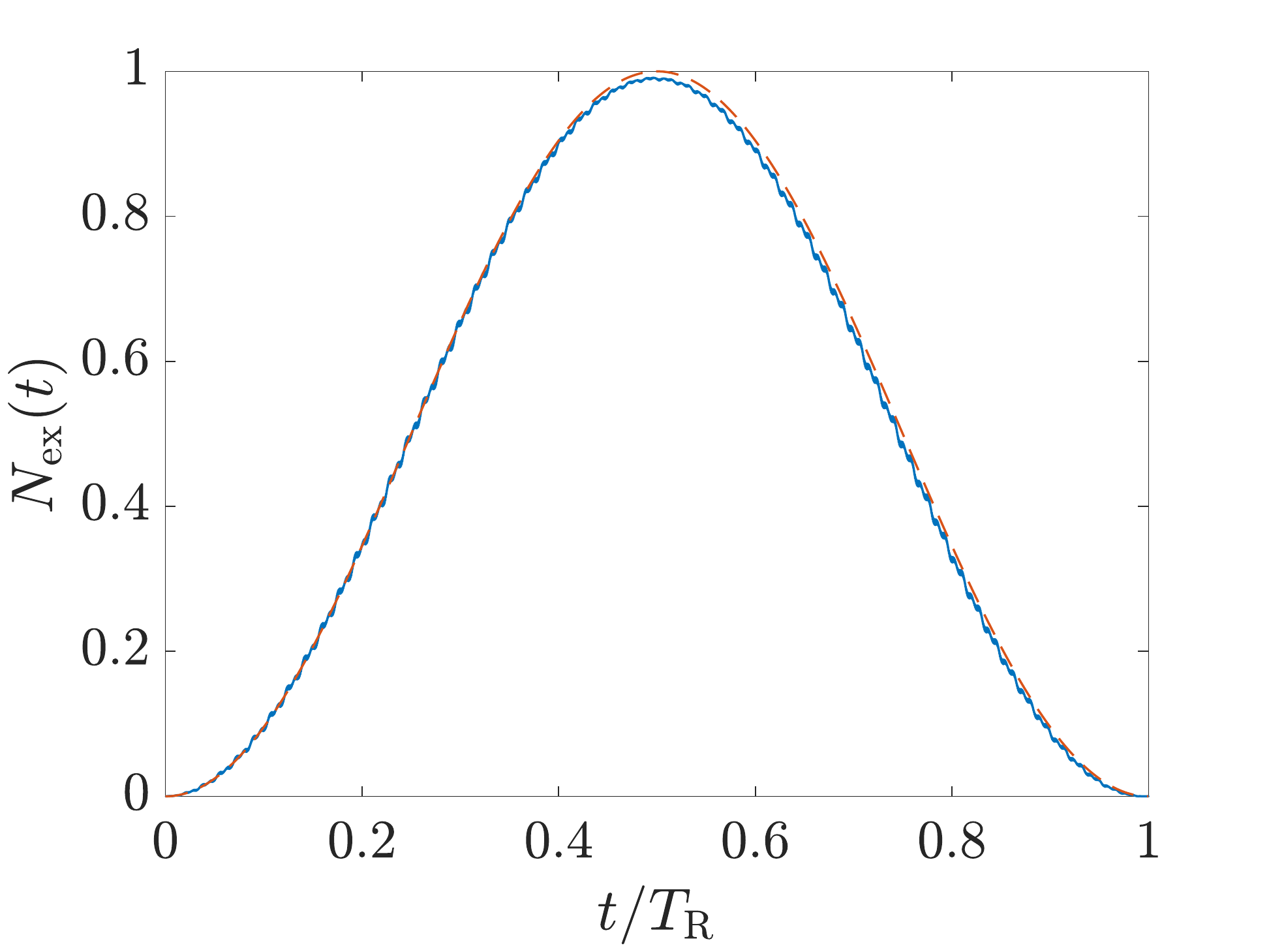}}
\subfigure[$\Delta = 10J$]{
\includegraphics[width= 0.23 \textwidth]{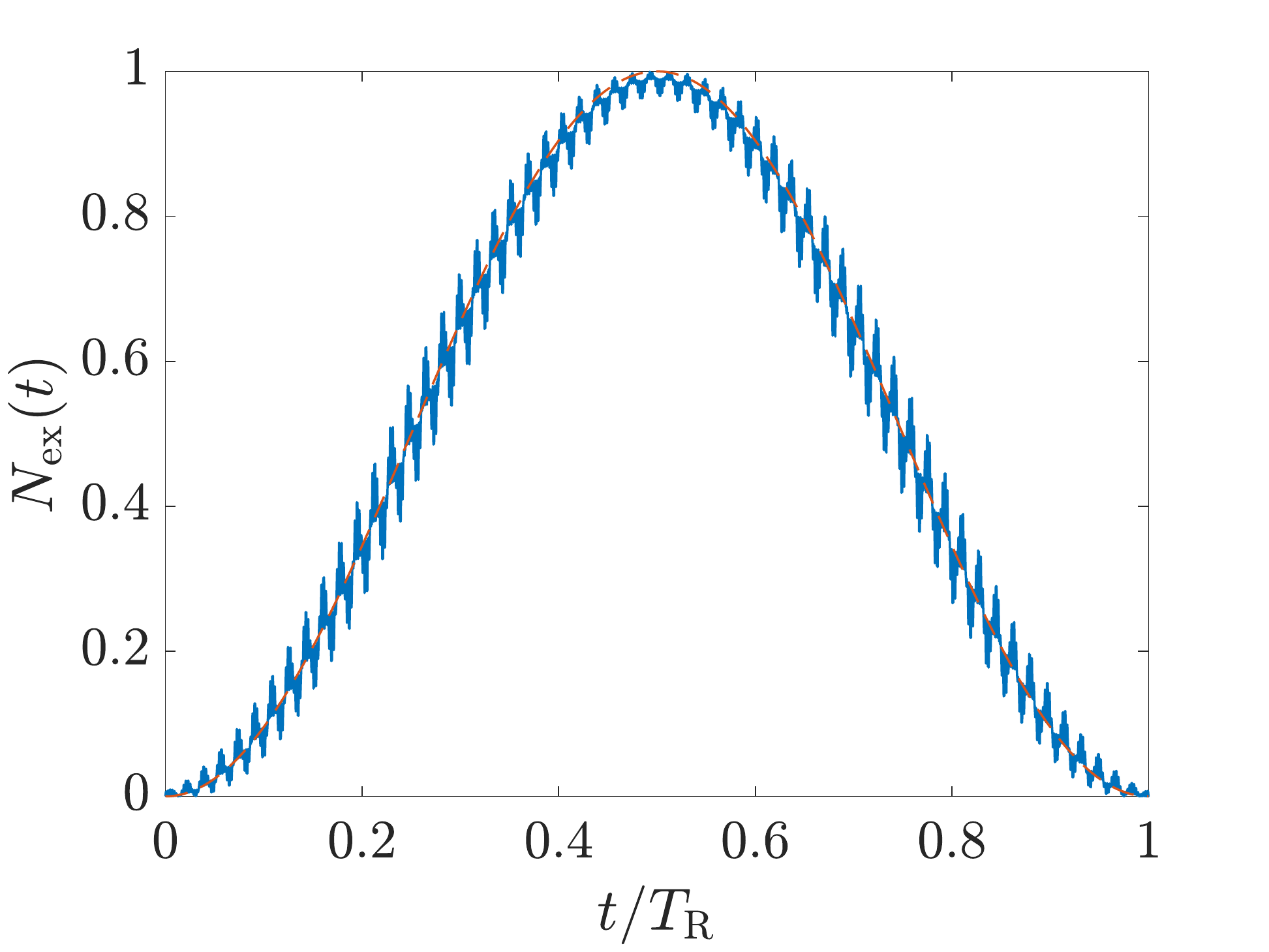}}
\subfigure[$\Delta = 3J$]{
\includegraphics[width= 0.23 \textwidth]{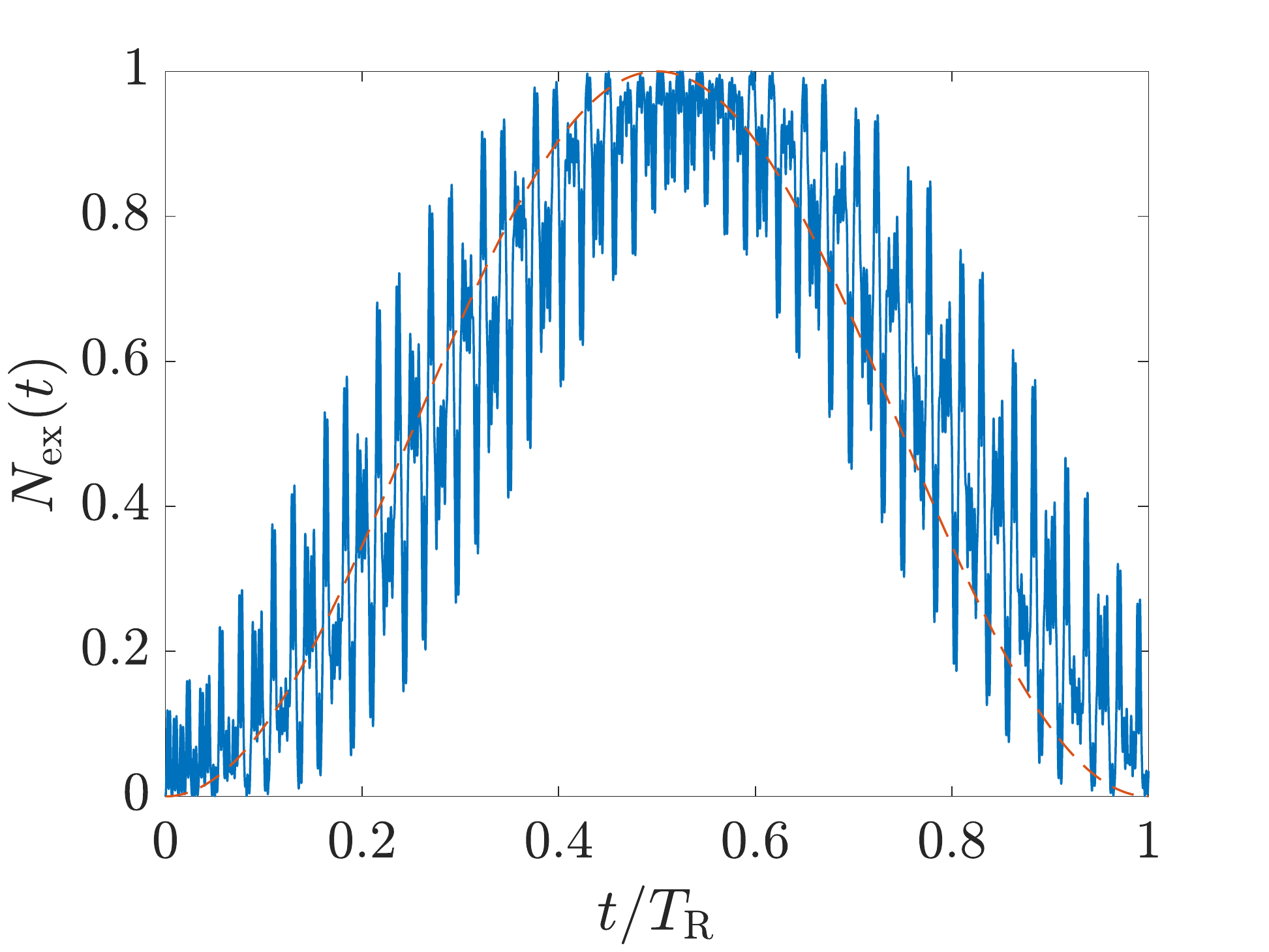}}
\subfigure[$\Delta = 2J$]{
\includegraphics[width= 0.23 \textwidth]{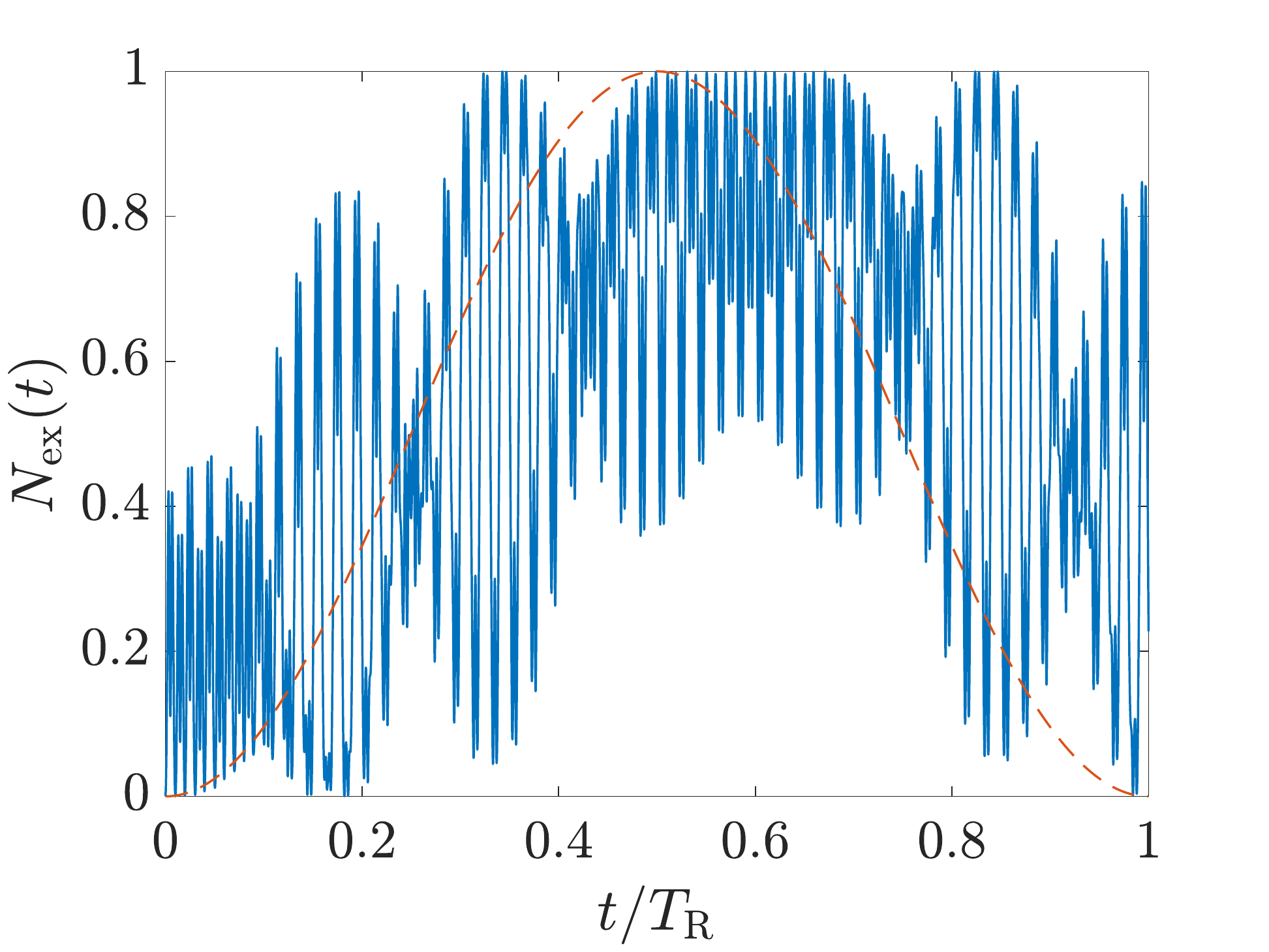}}
\caption{Excited fraction $N_{\text{ex}}(t)$, i.e.~the occupation in the high-energy dressed state over time, for various values of the initial detuning: $\Delta\!=\!100J$, $\Delta\!=\!10J$, $\Delta\!=\!3J$ and $\Delta\!=\!2J$. The thin red dotted line shows the perfect Rabi oscillation, as predicted by the simplified approach [Eq.~\eqref{simplification}], i.e. neglecting the micro-motion effects. The time-evolution shown by the blue curves is numerically generated from the full time-dependent Hamiltonians in Eqs.~\eqref{total_ham_frame}. The parameters are: $K_{\text{Fl}}\!=\!2\Delta$, $K_{\text{sp}}\!=\!0.01 J$ and the probe frequency $\omega_{\text{sp}}$ is adjusted so as to match the energy difference between the eigenstates of the exact effective Hamiltonian $\hat H_{\text{eff}}$ (as evaluated numerically).  Time is expressed in units of the Rabi period: $T_R\!=\!\pi/K_{\text{sp}}$.}
\label{Fig_Rabi}
\end{center}
\end{figure*}

\begin{figure}
\subfigure[$\Delta = 100J$]{
\includegraphics[width= 0.23 \textwidth]{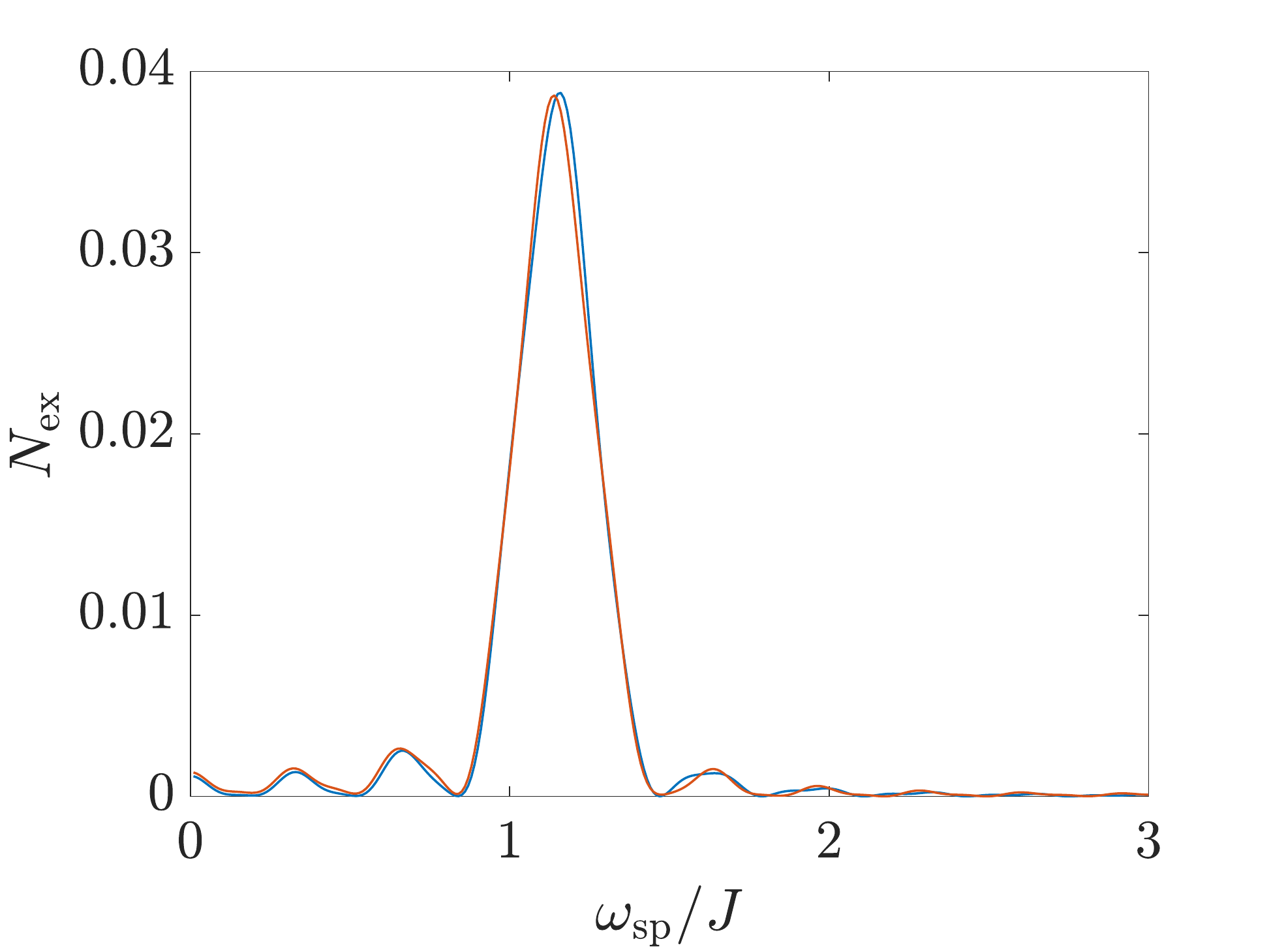}}
\subfigure[$\Delta = 10J$]{
\includegraphics[width= 0.23 \textwidth]{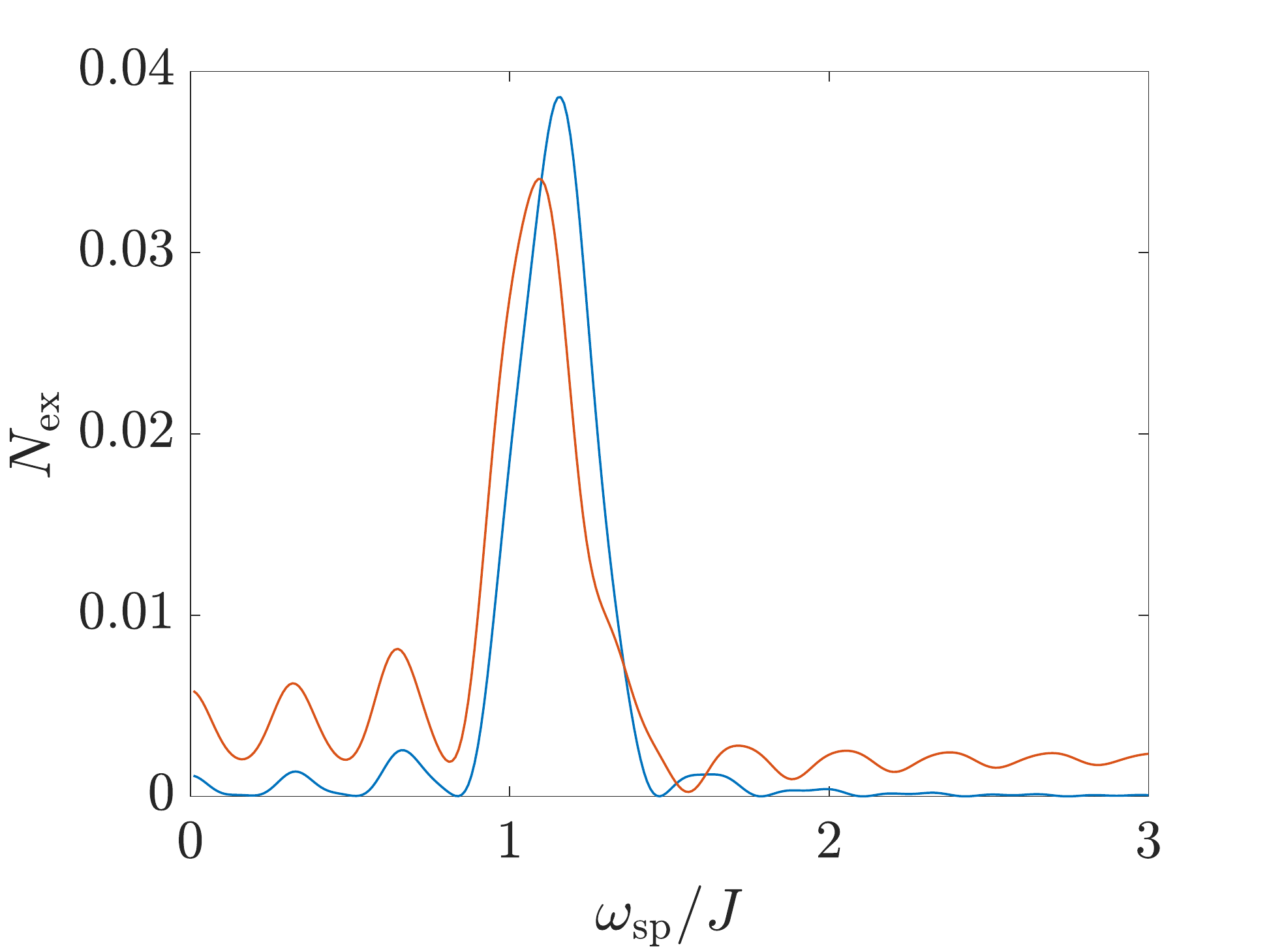}}
\caption{Excited fraction $N_{\text{ex}}(\omega_{\text{sp}})$ as a function of the probe frequency for $\Delta\!=\!100J$ and $\Delta\!=\!10J$. The spectra are obtained for stroboscopic (blue curve) and non-stroboscopic (red curves) observation times with respect to the Floquet period $T_{\text{Fl}}$. These times are chosen to be small compared to the Rabi period: $t^{\text{strob}}_{\text{obs}}\!=\!0.062T_{R}$ and $t^{\text{non-strob}}_{\text{obs}}\!=\!0.0637T_{R}$. Note that $t^{\text{strob}}_{\text{obs}}\!=\!0.062T_{R}$ corresponds to $31T_{\text{Fl}}$ (for $\Delta\!=\!10$) and $310T_{\text{Fl}}$ (for $\Delta\!=\!100$). 
}\label{Fig_Rabi_spectrum}
\end{figure}

We show in Fig.~\ref{Fig_Rabi} the excited fraction $N_{\text{ex}}(t)$, i.e.~the occupation in the high-energy dressed state, for various values of $\Delta$. Here, the dressed states $\vert +, - \rangle$ are obtained as the eigenstates of the exact effective Hamiltonian (evaluated numerically), and the probe frequency $\omega_{\text{sp}}$ is adjusted so as to match the energy difference between them (this allows one to avoid any residual detuning effect).  As shown in Fig.~\ref{Fig_Rabi}(a), the excited fraction depicts a perfect Rabi oscillation in the case of a large time-scale separation ($\Delta\!=\!100J$). As $\Delta$ is reduced, one observes a clear manifestation of the micro-motion effects [Fig.~\ref{Fig_Rabi}(b)-(c)]. In the absence of time-scale separation [Fig.~\ref{Fig_Rabi}(d)], the Rabi oscillation is barely recognizable. Next, we calculate the spectrum $N_{\text{ex}}(\omega_{\text{sp}})$, for a fixed probing time $t_{\text{obs}}$ that is small compared to the Rabi period, i.e.~$t_{\text{obs}}\!\ll\!1/K_{\text{sp}}$. In order to highlight the effects of micro-motion, we compare the spectra obtained for stroboscopic and non-stroboscopic times (with respect to the Floquet period $T_{\text{Fl}}$). The spectra shown in Fig.~\ref{Fig_Rabi_spectrum}(a) for $\Delta\!=\!100J$ reveal the well-know sinc-squared function associated with the Rabi formula~\cite{Cohen_book}, and the micro-motion effect is essentially invisible. This is no longer the case for the smaller value $\Delta\!=\!10J$, where the spectra obtained for stroboscopic and non-stroboscopic times are found to slightly differ. Finally, we test the relation in Eq.~\eqref{Rabi_integrate} by measuring the quantity
\be
\mathfrak{N}=(1/2 \pi K_{\text{sp}}^2)\int_{\omega_{\text{cut}}}^{\bar \omega_{\text{cut}}} \Gamma (\omega_{\text{sp}}) \text{d} \omega_{\text{sp}},\label{marker}
\ee
where the rate is obtained as $\Gamma (\omega_{\text{sp}})\!=\!N_{\text{ex}}(\omega_{\text{sp}})/t_{\text{obs}}$, and where we set the cut-offs at $\omega_{\text{cut}}\!=\!0.5 J$ and $\bar \omega_{\text{cut}}\!=\!2.5 J$; see Fig.~\ref{Fig_Rabi_spectrum}. From the spectra in Fig.~\ref{Fig_Rabi_spectrum}, we obtain
\begin{align}
&\Delta=100J: \quad \mathfrak{N}=0.96 \, \text{ for $t^{\text{strob}}_{\text{obs}}$}, \quad \mathfrak{N}=0.93 \, \text{ for $t^{\text{non-strob}}_{\text{obs}}$}, \notag \\
&\Delta=10J: \quad \mathfrak{N}=0.95 \, \text{ for $t^{\text{strob}}_{\text{obs}}$}, \quad \mathfrak{N}=1.13 \, \text{ for $t^{\text{non-strob}}_{\text{obs}}$},\notag 
\end{align}
where $t^{\text{strob}}_{\text{obs}}$ and $t^{\text{non-strob}}_{\text{obs}}$ refer to the stroboscopic and non-stroboscopic observation times. This shows that for $\Delta=10J$, an error of about 10\% can be attributed to the micro-motion; we note that this error can be reduced by optimizing the cut-offs.

\paragraph{Effect of time modulations with non-commuting operators}

In the previous subsection, we considered the case where the Floquet drive $\hat V_{\text{Fl}}(t)$ and the probe $\hat V_{\text{sp}} (t)$ commute with each other, which is also the case in the experiment described in the main text (where both drives correspond to a circular shake of the lattice). In this paragraph, we briefly discuss a subtlety that arises when treating a more generic situation where the two drives do not commute. 

To do so, we now modify the driven two-level system in Eq.~\eqref{total_ham} in a minimal manner by considering the time-dependent Hamiltonian
\begin{align} 
&\hat H (t)= \hat H_0 + \hat V_{\text{Fl}}(t) + \hat{\mathcal V}_{\text{sp}} (t), \label{total_ham_noncom}\\
&\hat H_0= \Delta \vert 1 \rangle \langle 1 \vert + J \left (\vert 0\rangle \langle 1\vert + \vert 1\rangle \langle 0\vert \right ),  \notag  \\
&\hat V_{\text{Fl}} (t)= K_{\text{Fl}} \cos (\Delta t) \vert 0 \rangle \langle 0 \vert , \notag \\
&\hat{\mathcal V}_{\text{sp}} (t)= 2K_{\text{sp}} \cos (\omega_S t) (-i |0\rangle \langle 1| + i |1\rangle \langle 0|) . \notag 
\end{align}
In contrast with the probe operator $\hat V_{\text{sp}} (t)$ in Eq.~\eqref{total_ham}, the two drives entering the problem no longer commute with each other: $[\hat V_{\text{Fl}},\hat{\mathcal V}_{\text{sp}}]\!\ne\!0$.

\begin{figure*}[!]
\begin{center}
\subfigure[$\Delta = 100J$]{
\includegraphics[width= 0.23 \textwidth]{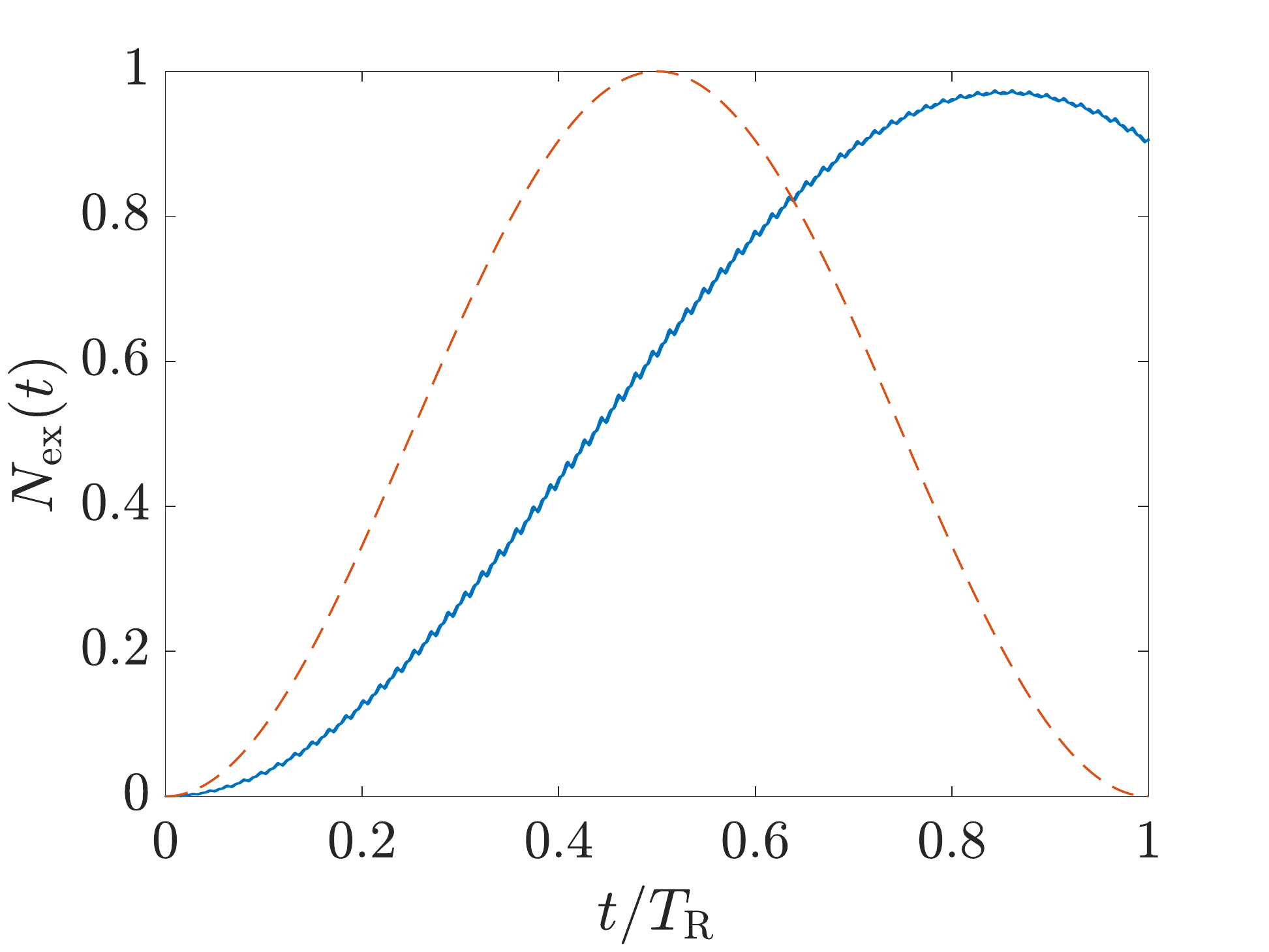}}
\subfigure[$\Delta = 10J$]{
\includegraphics[width= 0.23 \textwidth]{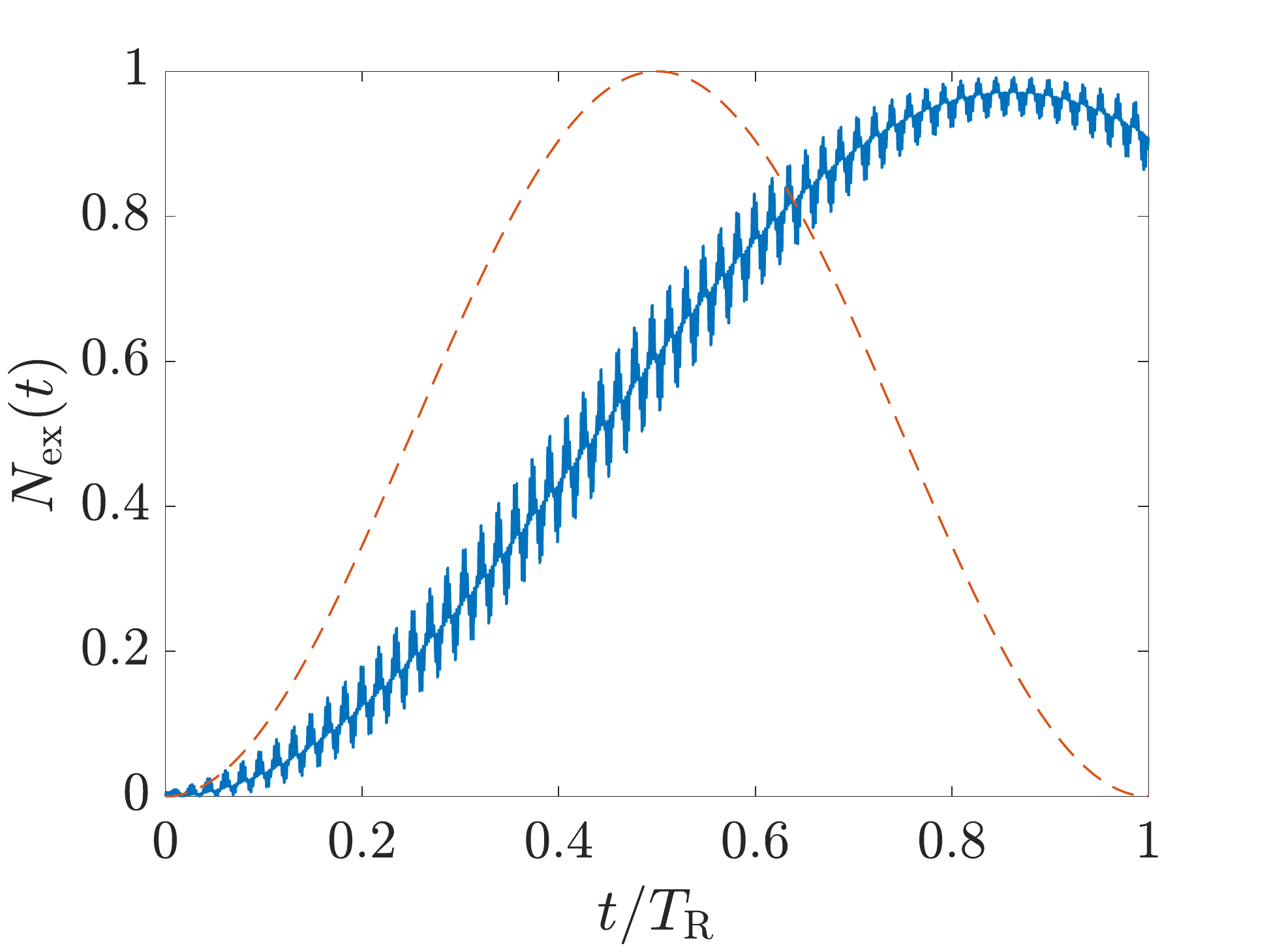}}
\subfigure[$\Delta = 3J$]{
\includegraphics[width= 0.23 \textwidth]{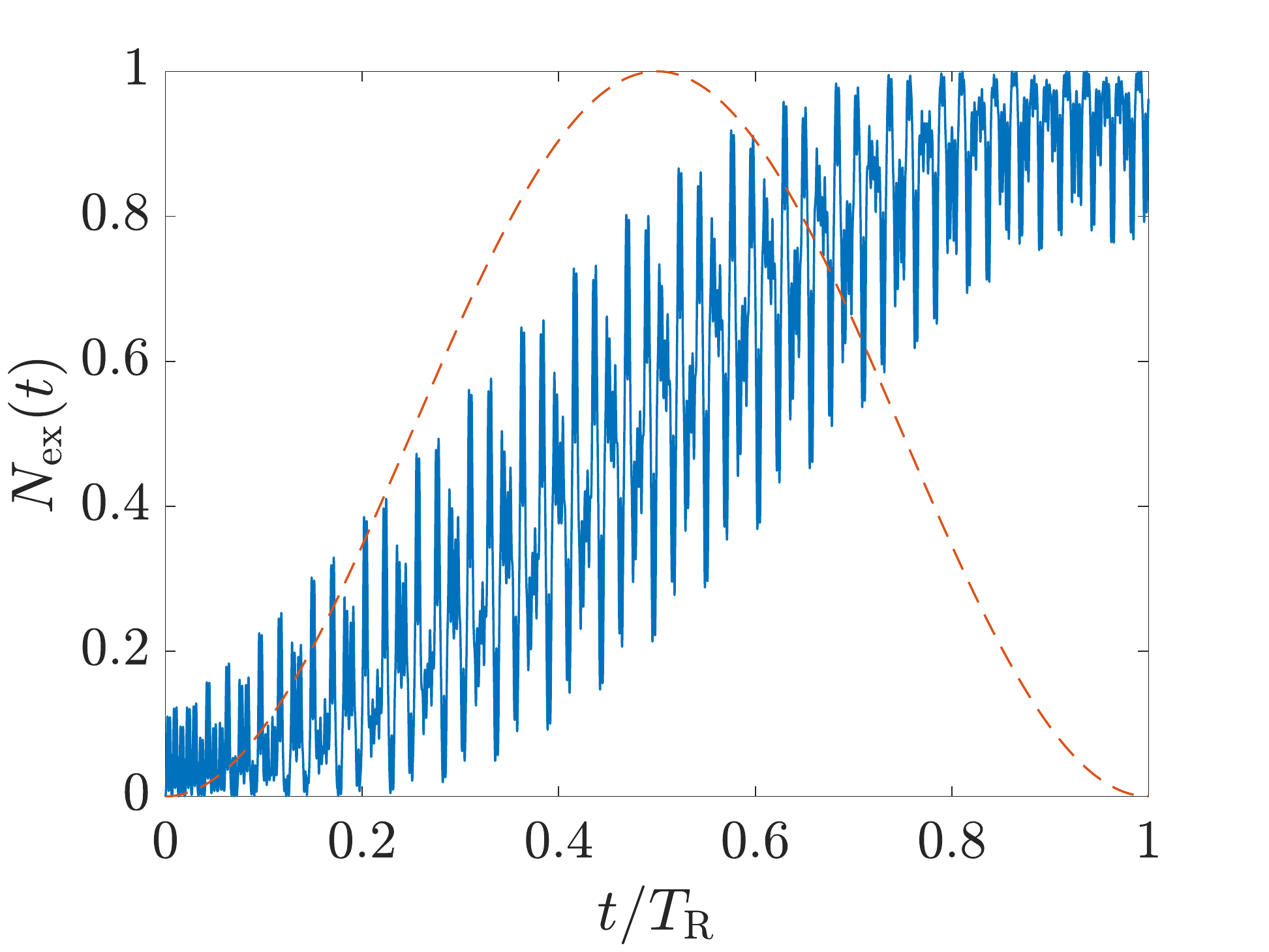}}
\subfigure[$\Delta = 2J$]{
\includegraphics[width= 0.23 \textwidth]{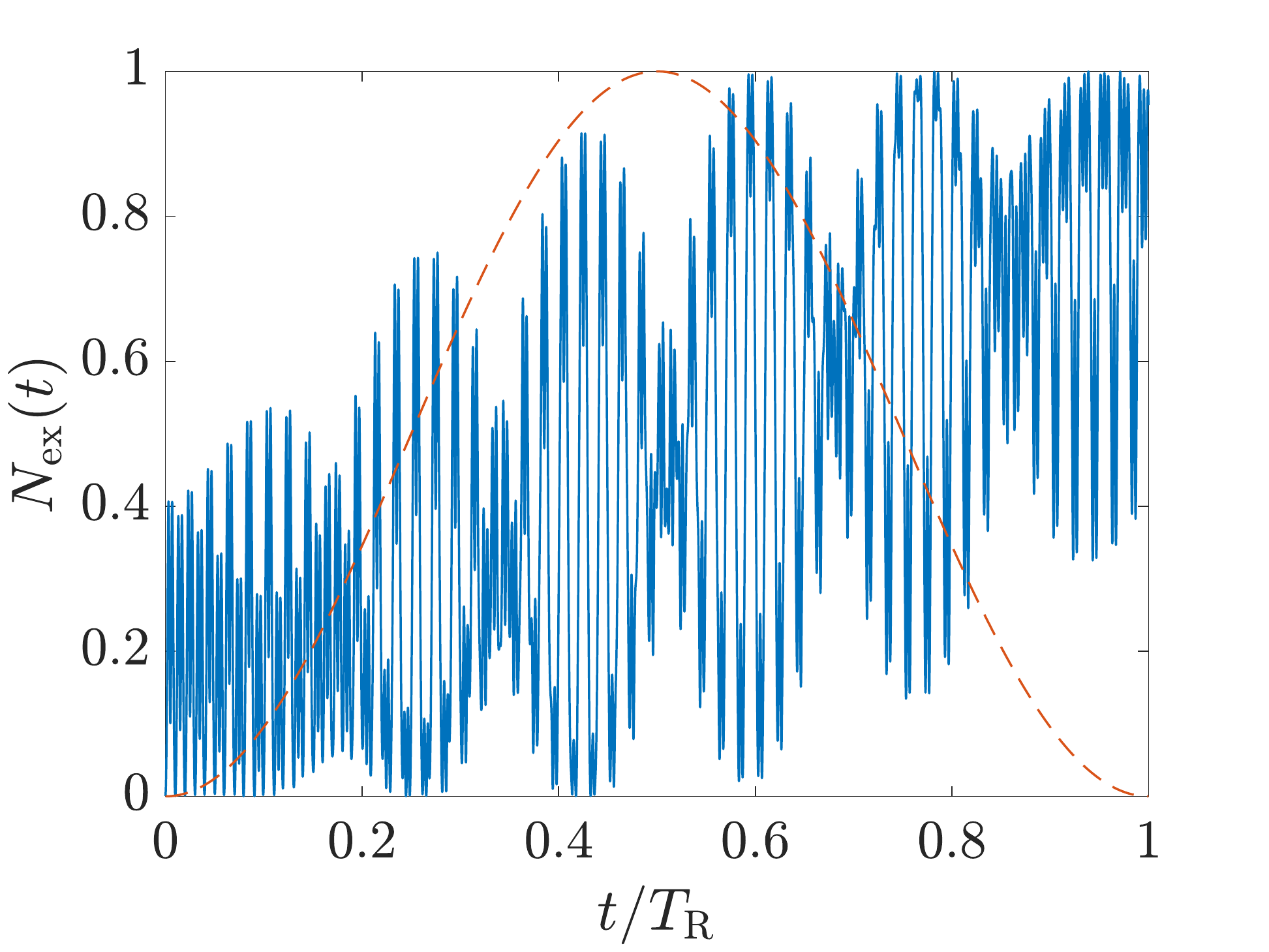}}
\caption{Excited fraction $N_{\text{ex}}(t)$, i.e.~the occupation in the high-energy dressed state over time, using the non-commutating probe $\hat {\mathcal V}_{\text{Sp}}(t)$ in Eq.~\eqref{total_ham_noncom}, and for various values of the initial detuning: $\Delta\!=\!100J$, $\Delta\!=\!10J$, $\Delta\!=\!3J$ and $\Delta\!=\!2J$. The thin red dotted line shows the Rabi oscillation, as predicted by the over-simplified approach [Eq.~\eqref{dressed_ham_noncom}], i.e. neglecting the non-commutativity effects. The time-evolution shown by the blue curves is numerically generated from the full time-dependent Hamiltonians in Eq.~\eqref{new_fram_noncom}, respectively. The parameters are: $K_{\text{Fl}}\!=\!2\Delta$, $K_{\text{sp}}\!=\!0.01 J$ and the probe frequency $\omega_{\text{sp}}$ is adjusted so as to match the energy difference between the eigenstates of the exact effective Hamiltonian $\hat H_{\text{eff}}$ (as evaluated numerically).  Time is expressed in units of the (naive) Rabi period: $T_R\!=\!\pi/K_{\text{sp}}$.}
\label{Fig_Rabi_noncom}
\end{center}
\end{figure*}

If one naively follows the simplified approach [Eq.~\eqref{simplification}], one recovers the analogue of Eq.~\eqref{dressed_ham},
 \be
\mathfrak{\hat H}_{\text{tot}} (t)=J \mathcal{J}_1 (K_{\text{Fl}}/\Delta) \hat \sigma_z - 2 K_{\text{sp}} \cos (\omega_{\text{sp}} t) \hat \sigma_y,\label{dressed_ham_noncom}
\ee
where we simply added the new probe $\hat{\mathcal V}_{\text{sp}} (t)$ on top of the effective Hamiltonian $\hat H_{\text{eff}}$ in Eq.~\eqref{eff_ham}, both expressed in the dressed-state basis. The expression in Eq.~\eqref{dressed_ham_noncom} justifies the choice of the new perturbation $\hat{\mathcal V}_{\text{sp}} (t)$ in Eq.~\eqref{total_ham_noncom}: Similarly to $\hat V_{\text{sp}} (t)$ in Eq.~\eqref{total_ham}, the probe $\hat{\mathcal V}_{\text{sp}} (t)$ is predicted to generate Rabi oscillations between the two dressed states upon setting the probe frequency $\omega_{\text{sp}}$ on resonance.

However, the above observation is not complete. Indeed, more care is required in this non-commutative context, as we now explain. As in the previous subsection, let us consider the frame transformation generated by the operator $\hat R (t)$ in Eq.~\eqref{frame_change}, and let us write the full time-dependent Hamiltonian~\eqref{total_ham_noncom} in this new frame,
\begin{align}
	&\hat{\mathcal{H}}(t)=\hat{R} \hat{H}(t) \hat{R}^\dagger - i \hat{R} \partial_t \hat{R}^\dagger \label{new_fram_noncom} \\
	&=J|0\rangle \langle 1| \exp \left[ i(K_{\text{Fl}}/\Delta) \sin (\Delta t) - i\Delta t \right] + \mathrm{h.c.} \notag \\
	&-2i K_{\text{sp}} \cos (\omega_{\text{sp}} t) |0\rangle \langle 1| 
	\exp \left[ i(K_{\text{Fl}}/\Delta) \sin (\Delta t) - i\Delta t \right]
	+\mathrm{h.c.}\notag
\end{align}
In the presence of time-scale separation $T_{\text{sp}}\!\gg\!T_{\text{Fl}}$, one can approximate the problem by performing a time-average of the Hamiltonian in Eq.~\eqref{new_fram_noncom} over a period $T_{\text{Fl}}$, while maintaining the time-dependence of the slow function $\cos (\omega_{\text{sp}} t)$. This more rigorous approach yields the effective time-dependent Hamiltonian
 \be
\mathfrak{\hat H}_{\text{tot}} (t)=\mathcal{J}_1(K_{\text{Fl}}/\Delta)  \left \{ J \hat \sigma_z - 2 K_{\text{sp}} \cos (\omega_{\text{sp}} t) \hat \sigma_y \right \},\label{dressed_ham_noncom_good}
\ee
which is similar to the more `naive' Hamiltonian in Eq.~\eqref{dressed_ham_noncom}, except that the coupling matrix elements are now renormalized by the Bessel function:  $K_{\text{sp}}\!\rightarrow\! \mathcal{J}_1(K_{\text{Fl}}/\Delta)K_{\text{sp}}$. Consequently, the Rabi frequency characterizing the resulting Rabi oscillations is reduced by a factor $\mathcal{J}_1 (K_{\text{Fl}}/\Delta)$, with respect to the case of the commutating probe $\hat V_{\text{sp}}$. This observation is subtle, but crucial, whenever one probes a system with a non-commutating probe operator in view of extracting the coupling matrix elements entering Eq.~\eqref{Rabi_integrate}, e.g.~to extract the Chern number~\cite{Tran2017, Tran2018}.

We validate this result in Fig.~\ref{Fig_Rabi_noncom}, where we plot the analogue of Fig.~\ref{Fig_Rabi} now for the non-commutating probe $\hat{\mathcal V}_{\text{sp}}$. Here, one clearly visualizes a striking elongation of the Rabi period $T_R$ corresponding to the factor $1/\mathcal{J}_1 (K_{\text{Fl}}/\Delta)\!\approx\!1.73$, as predicted above. Besides, we note that the micro-motion effects that appear as one decreases the value of $\Delta$ are slightly enhanced as compared to the commutative case [Fig.~\ref{Fig_Rabi}]. Most importantly, this renormalization of the Rabi frequency strongly modifies the integrated rate relation in Eq.~\eqref{Rabi_integrate}. To see this, we now calculate the following quantity
\be
\mathfrak{N}_{\text{renorm}}=[1/2 \pi (\mathcal{J}_1(K_{\text{Fl}}/\Delta)K_{\text{sp}})^2]\int_{\omega_{\text{cut}}}^{\bar \omega_{\text{cut}}} \Gamma (\omega_{\text{sp}}) \text{d} \omega_{\text{sp}},
\ee
which differs from the marker in Eq.~\eqref{marker} in that $\mathfrak{N}_{\text{renorm}}$  explicitly takes into account the renormalized Rabi frequency, i.e.~$K_{\text{sp}}\!\rightarrow\! \mathcal{J}_1(K_{\text{Fl}}/\Delta)K_{\text{sp}}$. From our numerical simulations, we obtain  $\mathfrak{N}_{\text{renorm}}\!=\!1.02$ for $\Delta\!=\!100J$ and $\mathfrak{N}_{\text{renorm}}\!=\!1.20$ for $\Delta\!=\!10J$ (using the same parameters as above). This numerical analysis illustrates how probing Floquet systems should be treated with care, when using spectroscopic probes that do not commute with the Floquet drive. We point out that while these effects do not affect the experiment discussed in the main text (where both drives commute with each other), we anticipate that they might play a role in other Floquet-based experimental settings.

\end{document}